\documentclass[english,12pt,draftclsnofoot,english,onecolumn]{IEEEtran}
\usepackage[T1]{fontenc}
\usepackage{color}
\usepackage{babel}
\usepackage{amsmath}
\usepackage{amssymb}
\usepackage{graphicx}
\usepackage{setspace}
\usepackage[unicode=true,
 bookmarks=true,bookmarksnumbered=true,bookmarksopen=true,bookmarksopenlevel=1,
 breaklinks=false,pdfborder={0 0 0},pdfborderstyle={},backref=false,colorlinks=false]
 {hyperref}
\hypersetup{pdftitle={Your Title},
 pdfauthor={Your Name},
 pdfpagelayout=OneColumn, pdfnewwindow=true, pdfstartview=XYZ, plainpages=false}

\makeatletter

\providecommand{\tabularnewline}{\\}

\ifCLASSOPTIONcompsoc
\usepackage[caption=false,font=normalsize,labelfont=sf,textfont=sf]{subfig}
\else
\usepackage[caption=false,font=footnotesize]{subfig}
\fi
\usepackage{cite}
\usepackage{amsthm}
\newtheorem{theorem}{\textbf{Theorem}}
\newtheorem{corollary}{\textbf{Corollary}}
\newtheorem{lemma}{\textbf{Lemma}}
\newtheorem{proposition}{\textbf{Proposition}}

\usepackage{algorithm}
\usepackage{amsmath}

\@ifundefined{showcaptionsetup}{}{%
 \PassOptionsToPackage{caption=false}{subfig}}
\usepackage{subfig}
\makeatother

\begin{document}
\title{Access Points in the Air: Modeling and Optimization of Fixed-Wing
UAV Network}
\author{\IEEEauthorblockN{Junyu Liu, Min Sheng, Ruiling Lyu, Yan Shi, Jiandong
Li}\\
\IEEEauthorblockA{State Key Laboratory of ISN, Xidian University,
Xi'an, Shaanxi, 710071, China}}
\maketitle
\begin{abstract}
\textcolor{black}{Fixed-wing unmanned aerial vehicles (UAVs) are of
great potential to serve as aerial access points (APs) owing to better
aerodynamic performance and longer flight endurance. However, the
inherent hovering feature of fixed-wing UAVs may result in discontinuity
of connections and frequent handover of ground users (GUs). In this
work, we model and evaluate the performance of a fixed-wing UAV network,
where UAV APs provide coverage to GUs with millimeter wave backhaul.
Firstly, it reveals that network spatial throughput (ST) is independent
of the hover radius under real-time closest-UAV association, while
linearly decreases with the hover radius if GUs are associated with
the UAVs, whose hover center is the closest. Secondly, network ST
is shown to be greatly degraded with the over-deployment of UAV APs
due to the growing air-to-ground interference under excessive overlap
of UAV cells. Finally, aiming to alleviate the interference, a projection
area equivalence (PAE) rule is designed to tune the UAV beamwidth.
Especially, network ST can be sustainably increased with growing UAV
density and independent of UAV flight altitude if UAV beamwidth inversely
grows with the square of UAV density under PAE.}
\end{abstract}

\section{Introduction\label{sec:Introduction}}

Benefiting from rapid event response and flexible deployment, unmanned
aerial vehicle (UAV) networks have experienced explosive development
and evolution in recent years \cite{Ref_background_1,Ref_backhaul_Yanjie}.
Among the applications of great potential, low-altitude UAVs are capable
of serving as the access points (APs) to provide service to ground
users (GUs) in short-distance line-of-sight wireless channels \cite{Ref_background_2,Ref_rotor_UAV_5}.
Moreover, the deployment of UAV APs is flexible to provide instant
coverage in a variety of scenarios including public safety, dense
crowds, internet of things (IoT) applications and emergency scenarios.
Thanks to better aerodynamic performance, greater payload versatility
and longer flight endurance, fixed-wing UAVs are promising in serving
as the APs in the air \cite{Ref_background_fixed_1}. Moreover, millimeter
wave (mmWave) by nature can be used to convey wireless backhaul for
UAV APs, since there are few obstacles for the air-to-ground channel.
In particular, it was reported that up to 10 Gbps peak rates could
be reached using multi-user multiple-input-multiple-output (MU-MIMO)
mmWave for long-range transmissions \cite{Ref_Background_mmWave_1}.

Despite the great potential, fixed-wing UAVs have to hover in the
air when providing coverage to a given ground area. With the increase
of hover radius, it is difficult for fixed-wing UAVs to provide steady
and continuous service to GUs. Moreover, mmWave beams are basically
oriented to the hovering UAVs to guarantee the backhaul capacity.
Therefore, increasing the UAV hover radius may potentially enlarge
the mmWave beamwidth, thereby degrading the mainlobe gain and backhaul
capacity. Hence, it is crucial to investigate the impact of UAV hover
radius and how to effectively enhance the performance of the fixed-wing
UAV network.

\subsection{Related Work}

The research on UAV networks, where UAVs serve as APs, has attracting
considerable attention from both academia and industry \cite{Ref_rotor_UAV_1,Ref_rotor_UAV_2,Ref_rotor_UAV_3,Ref_rotor_UAV_4,Ref_Spectrum_Sharing,Ref_rotor_UAV_5}.
The UAV trajectory optimization problem was considered in \cite{Ref_rotor_UAV_2},
where UAVs provided service to cell edge users to offload the data
traffic of terrestrial cellular network. Specifically, UAV trajectory
and user scheduling were alternately optimized to maximize the sum
rate of UAV-served edge users with the rate constraint of all the
users. In \cite{Ref_rotor_UAV_3}, a distributed UAV deployment algorithm
was proposed to minimize the distance from UAV APs to GUs so as to
improve the user coverage. Moreover, the UAV AP deployment problem
has been extended into heterogeneous networks \cite{Ref_rotor_UAV_4}.
Especially, a latency-aware approach has been proposed, which jointly
optimizes the location and association coverage of UAVs, to minimize
the total average latency of GUs considering the existing terrestrial
base stations (TBSs). If frequency resources are shared by UAV APs
and TBSs, potential cross-layer interference will be generated, which
may degrade the performance of the UAV network. The performance of
a UAV integrated terrestrial cellular network is evaluated in \cite{Ref_Spectrum_Sharing,Ref_rotor_UAV_5}.
Especially, the impact of UAV deployment density on the performance
of the coexisting system was captured in \cite{Ref_Spectrum_Sharing}
and an effective interference avoidance scheme was proposed to mitigate
the air-to-ground interference and significantly improve the spectrum
efficiency of the coexisting system in \cite{Ref_rotor_UAV_5}.

Due to flexible deployment and low expenditure, rotary-wing UAVs are
applied in most of the available research \cite{Ref_rotor_UAV_1,Ref_rotor_UAV_2,Ref_rotor_UAV_3,Ref_rotor_UAV_4,Ref_Spectrum_Sharing,Ref_rotor_UAV_5},
where rotary-wing UAVs were assumed to move according to the preset
trajectory and steadily stay in a given position. Accordingly, rotary-wing
UAV systems have been designed and developed by a number of companies
and operators \cite{Ref_rotor_UAV_KDDI,Ref_rotor_UAV_ATT}. However,
since most of off-the-shelf rotary-wing UAVs are battery driven, they
are energy-inefficient and excessive power is required especially
when they are taking off, climbing and changing the flight attitude.
Therefore, the flight endurance of the rotary-wing UAVs is significantly
limited (dozens of minutes for battery-driven ones). Moreover, limited
wireless backhaul will bottleneck the performance of the UAV system
supposing that high data rates are required by the connected GUs.
To address the issues of limited flight endurance and backhaul, AT\&A
was reported to develop a tethered UAV system, named Flying COW \cite{Ref_rotor_UAV_ATT}.
Particularly, each rotary-wing UAV, which was connected to the ground
by a thin tether, is capable of providing the long-term evolution
(LTE) coverage to GUs. The tether was used for conveying highly secure
backhaul via fiber and supplying power, which allows for longer flight
time.

Although Flying COW is designed to ultimately provide coverage to
an area up to 40 square miles, the tether connection will inevitably
limit the deployment and mobility of UAVs. On the other hand, researchers
and engineers are increasingly focusing on the development of the
fixed-wing UAVs \cite{Ref_background_fixed_3,Ref_background_fixed_4,Ref_background_fixed_1,Ref_background_fixed_2}.
Compared with the rotary-wing UAVs, the deployment of fixed-wing UAVs
is more flexible. For instance, fixed-wing UAVs could offer better
aerodynamic performance, which makes them well suited for the application
in higher flight altitude, greater payload and longer ranges and flight
endurance. In \cite{Ref_background_fixed_2}, an all-weather fixed-wing
emergency communication system was developed for the quick network
recovery in the emergency communication scenarios. According to the
aerodynamic principle, nevertheless, fixed-wing UAVs cannot steadily
stay in a given position and have to hover in the air. If the hover
radius is large, rotary-wing UAVs would fail to provide continuous
service to GUs. Moreover, although sufficient backhaul can be provided
through mmWave link \cite{Ref_mmWave_channel_RSH_2018,Ref_mmWave_channel_RSH_2020},
the high mobility of the hovering fixed-wing UAVs may seriously deteriorate
mmWave link capacity since the mmWave beam is highly directional.
Worse still, the acquisition of real-time perfect channel state information
(CSI) in the highly dynamic network is challenging and remains to
be an open problem \cite{Ref_perfect_CSI}. Especially, the acquired
CSI would be easily outdated at the moment of decision due to the
on-the-move feature of fixed-wing UAVs. Therefore, the performance
and deployment of fixed-wing UAV network remains to be further investigated,
which motivates this work.

\subsection{Contribution and Outcome}

In this paper, we model a downlink UAV network, where fixed-wing UAV
APs provide service to GUs and mmWave is applied to provide wireless
backhaul. In particular, the impact of key parameters, including UAV
flight altitude, deployment density, hover radius, backhaul limitation
and user association rules, etc., on the performance of UAV network
has been evaluated. On this basis, we further investigate how to effectively
improve the UAV network performance through adjusting UAV beamwidth.
The main conclusions of this work are summarized as follows. 
\begin{itemize}
\item \textbf{Real-time VS semi-real-time user association rules.} We evaluate
the performance of two typical user association rules, namely, real-time
user association (RTNA), where GUs always connect the closest UAVs,
and semi-real-time user association (Semi-RTNA), where GUs connect
to the UAVs whose hover centers are the closest, in terms of network
spatial throughput (ST). It is shown that network ST is greatly degraded
by increasing the UAV hover radius under Semi-RTNA. The reason is
that desired signal power is likely to be reduced if the associated
UAVs hover apart under large hover radius. Worse still, the increase
of hover radius would notably increase the mmWave mainlobe beamwidth,
thereby reducing the mainlobe gain and backhaul capacity. On the contrary,
the impact of hover radius on the performance of RTNA is shown to
be minor even though more handover overhead is introduced. Especially,
if ignoring the backhaul limitation, network ST is shown to be independent
of the hover radius under RTNA. 
\item \textbf{Optimization of UAV beamwidth.} The impact of UAV beamwidth
on network ST is further investigated. Particularly, reducing the
UAV beamwidth first increases (due to alleviating overlap-cell interference)
and then decreases (due to limiting UAV coverage) network ST. Moreover,
the optimal beamwidth is shown to be dependent on the activated UAV
density $\lambda_{\mathrm{a}}$. Accordingly, we propose a projection
area equivalence policy to optimize the UAV beamwidth. In particular,
network ST could be sustainably increased with UAV density if UAV
half-beamwidth inversely grows with $\frac{1}{\sqrt{\lambda_{\mathrm{a}}}}$.
More importantly, the optimized network ST is proved to be independent
of the UAV flight altitude in the backhaul-unlimited case. In other
words, aerial spatial resources can be fully exploited by the proposed
policy.
\end{itemize}

We organize the remaining parts of this paper as follows. System model
is given in Section \ref{sec:System-Model} followed by the performance
analysis of two typical user association rules in UAV network with
respect to user coverage probability and network ST in Section \ref{sec:Analysis}.
In Section \ref{sec:Optimization}, we further investigate the impact
of UAV antenna beamwidth on the performance of directional-antenna
UAV network and optimize network ST through adjusting UAV beamwidth.
Finally, conclusions are drawn in Section \ref{sec:Conclusion}. The
main parameter notations used in the paper are summarized in Table
\ref{tab:Summary-of-Notation}.

\begin{table*}[t]
\centering{}\caption{\label{tab:Summary-of-Notation}Summary of Notations}
\begin{tabular}{|c|c|c|c|}
\hline 
Symbol & Meaning & Symbol & Meaning\tabularnewline
\hline 
\hline 
$\mathrm{UAV}_{i}$, $\mathrm{GU}_{j}$ & $i$th TBS, $j$th UAV & \multicolumn{1}{c|}{$\Phi$} & UAV half-beamwidth\tabularnewline
\hline 
$\Pi_{\mathrm{UAV}}$, $\Pi_{\mathrm{GU}}$ & UAV and GU sets & $G\left(\phi,\varphi\right)$ & antenna mainlobe gain\tabularnewline
\hline 
$\mathrm{\tilde{\Pi}_{UAV}}$ & activated UAV set & $R_{\mathrm{p}}$ & projection radius\tabularnewline
\hline 
$\lambda$, $\lambda_{\mathrm{GU}}$ & UAV and GU densities & \multicolumn{1}{c|}{$p_{\mathrm{p}}$} & projection probability\tabularnewline
\hline 
$\lambda_{\mathrm{a}}$ & activated UAV density & $\alpha$ & pathloss exponent\tabularnewline
\hline 
$q_{\mathrm{a}}$ & UAV activated probability & $R_{\mathrm{m}}$ & backhaul coverage radius\tabularnewline
\hline 
$\eta$ & $\eta=\lambda/\lambda_{\mathrm{GU}}$ & $C_{\mathrm{t}}$ & normalized backhaul capacity\tabularnewline
\hline 
$\mu$ & $\mu$=3.5 & $C_{\mathrm{b}}$ & backhaul capacity for each UAV\tabularnewline
\hline 
$R_{\mathrm{h}}$ & UAV hover radius & $\vartheta$ & mmWave mainlobe beamwidth\tabularnewline
\hline 
$\theta$ & UAV hover angle & $G_{\mathrm{m}}\left(\vartheta\right)$ & mmWave mainlobe gain\tabularnewline
\hline 
$h_{\mathrm{UAV}}$, $h_{\mathrm{GU}}$ & UAV and GU altitudes & $\varepsilon$ & mmWave orientation error\tabularnewline
\hline 
$h_{\mathrm{L}}$ & lower bound of $h_{\mathrm{UAV}}$ & $\bar{\varepsilon}$ & mean of $\varepsilon$\tabularnewline
\hline 
$h_{\mathrm{U}}$ & upper bound of $h_{\mathrm{UAV}}$ & $F_{\left|\varepsilon\right|}\left(\vartheta\right)$ & mmWave orientation error\tabularnewline
\hline 
$\Delta h$ & $\Delta h=h_{\mathrm{UAV}}-h_{\mathrm{GU}}$ & $r_{0}$ & 2D distance from $\mathrm{UAV}_{0}$ to $\mathrm{GU}_{0}$\tabularnewline
\hline 
$\tilde{h}_{\mathrm{L}}$ & \textcolor{black}{$\tilde{h}_{\mathrm{L}}=h_{\mathrm{L}}-h_{\mathrm{GU}}$} & $C$ & projection scaling parameter\tabularnewline
\hline 
$\tilde{h}_{\mathrm{U}}$ & \textcolor{black}{$\tilde{h}_{\mathrm{U}}=h_{\mathrm{U}}-h_{\mathrm{GU}}$} & $P$ & UAV transmit power\tabularnewline
\hline 
\end{tabular}
\end{table*}

\section{System Model\label{sec:System-Model}}

\subsection{Network Model\label{subsec:Network Model}}

Consider a downlink UAV network, where fixed-wing UAV APs provide
service to GUs. For the two-dimension (2D) locations, UAVs and GUs
are distributed as two independent homogeneous Poisson Point Processes
(PPPs) $\Pi_{\mathrm{UAV}}=\left\{ \mathrm{UAV}_{i}\right\} $ $\left(i\in\mathbb{N}\right)$
with density $\lambda$ and $\Pi_{\mathrm{GU}}=\left\{ \mathrm{GU}_{j}\right\} $
$\left(j\in\mathbb{N}\right)$ with density $\lambda_{\mathrm{GU}}$,
respectively. \textcolor{black}{For practical concern, we assume that
fixed-wing UAVs hover in the air with identical hover radius $R_{\mathrm{h}}$
and different flight altitude $h_{\mathrm{UAV}}$, which follows uniform
distribution $h_{\mathrm{UAV}}\sim\mathcal{U}\left(h_{\mathrm{L}},h_{\mathrm{U}}\right)$.
Supposing that GUs are of identical height $h_{\mathrm{GU}}$, the
vertical distance $\Delta h\left(=h_{\mathrm{UAV}}-h_{\mathrm{GU}}\right)$
from UAVs to GUs follows uniform distribution $\Delta h\sim\mathcal{U}\left(\tilde{h}_{\mathrm{L}},\tilde{h}_{\mathrm{U}}\right)$,
where $\tilde{h}_{\mathrm{L}}=h_{\mathrm{L}}-h_{\mathrm{GU}}$ and
$\tilde{h}_{\mathrm{U}}=h_{\mathrm{U}}-h_{\mathrm{GU}}$. }It is assumed
that multiple GUs can simultaneously connect to one UAV AP for service.
For scheduling fairness, each UAV AP randomly and independently serves
the connected GUs in a time-division manner. Therefore, the GUs in
one UAV cell have the equal chance to be served. To improve frequency
reuse, spectrum is reused by different UAV cells. As a result, the
neighboring UAV APs may potentially generate inter-cell interference
to the intended GU. If not properly handled, the inter-cell interference
may significantly degrade the performance of the downlink UAV network.
Besides, saturated data model is adopted such that each GU always
has data request from the connected UAVs.

\begin{figure}[t]
\begin{centering}
\includegraphics[width=3.5in]{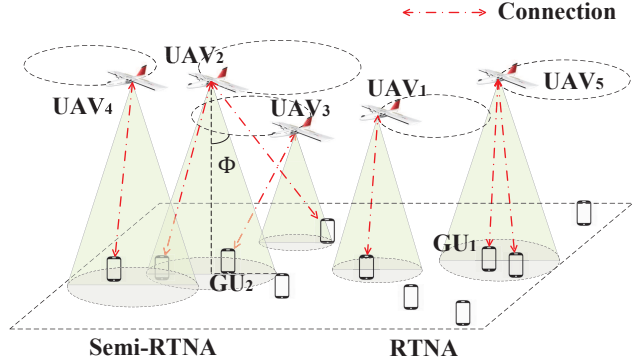}
\par\end{centering}
\caption{\label{fig:system model}Illustration of a fixed-wing UAV network.
RTNA and Semi-RTNA are used for user association.}

\end{figure}

\subsection{User Association Model}

We adopt the following two user association rules to balance the UAV
coverage performance and handover overhead.
\begin{itemize}
\item Real-time nearest association (RTNA). One GU is associated with the
UAV, which is horizontally closest to the GU\footnote{The flight altitude of UAV APs may be varying due to air buoyancy
and attitude control to ensure that the hovering UAVs could cover
a given area. Therefore, each GU is assumed to connect to horizontally
closest UAV for stronger average received power and better coverage.}. It is shown in Fig. \ref{fig:system model} that $\mathrm{GU}_{1}$
is located closer to the hover center of $\mathrm{UAV}_{1}$ compared
to that of $\mathrm{UAV}_{5}$. However, the hovering $\mathrm{UAV}_{5}$
is closer to $\mathrm{GU}_{1}$ in the current time slot. Therefore,
$\mathrm{GU}_{1}$ is associated with $\mathrm{UAV}_{5}$ under RTNA.
In the coming time slots, when $\mathrm{UAV}_{1}$ hovers closer,
$\mathrm{GU}_{1}$ will be handover to $\mathrm{UAV}_{1}$.
\item Semi-real-time nearest association (Semi-RTNA). One GU is associated
with the UAV, the hover center of which is horizontally closest to
the GU. It is shown in Fig. \ref{fig:system model} that, even if
$\mathrm{GU}_{2}$ is closer to $\mathrm{UAV}_{2}$ in the current
time slot, $\mathrm{GU}_{2}$ is associated with $\mathrm{UAV}_{3}$
under Semi-RTNA to reduce the handover overhead.
\end{itemize}

It is assumed that real-time location information of UAVs could be
acquired by GUs for performing RTNA and Semi-RTNA. Accordingly, GUs
could calculate the distance to the UAVs, which could provide coverage,
to determine the UAV to connect to. Then, the handover of user association
is initiated by the GU, which sends the handover request to the associated
UAV AP and the UAV AP to connect to. Note that the handover frequency
is dependent on the location of GUs, number of fixed-wing UAV APs
near the GUs, hover radius and velocity of UAV APs. If not properly
handled, the frequent handover will result in considerable handover
overhead, thereby degrading the performance of the UAV network.

\subsection{Antenna and Channel Model\label{subsec:Channel Model}}

It is shown in Fig. \ref{fig:system model} that each UAV with constant
transmit power $P$ is equipped with one directional antenna. The
azimuth and elevation beamwidths are $2\Phi$ with $\Phi\in\left(0,\frac{\pi}{2}\right)$.
Denoting $\phi$ and $\varphi$ as the azimuth and elevation angles,
respectively, the antenna gain in the direction $\left(\phi,\varphi\right)$
(corresponding to the antenna mainlobe) is modeled as \cite{Ref_UAV_beam_model}
\begin{equation}
G\left(\phi,\varphi\right)=\begin{cases}
G_{\mathrm{0}}/\Phi^{2}, & -\Phi\leq\phi\leq\Phi,-\Phi\leq\varphi\leq\Phi,\\
0, & \mathrm{otherwise},
\end{cases}\label{eq:antenna gain}
\end{equation}
where $G_{0}=2.2846$. Accordingly, the projection of each UAV on
the ground is a disk region with radius $R_{\mathrm{p}}=\Delta h\tan\Phi$.
If one GU is out of the projection area of any UAV, no service will
be provided. Note that, although increasing $\Phi$ could enhance
the UAV coverage, it will result in greater overlap of the projection
areas, which may introduce excessive inter-cell interference.

The channel from UAVs to GUs is assumed to consist of pathloss $d^{-\alpha}$
and Rayleigh fading $h\sim\mathcal{CN}\left(0,1\right)$ \cite{Ref_OFDMA_1,Ref_OFDMA_2},
where $d$ and $\alpha$ denote the distance from one UAV to its associated
GU and pathloss exponent, respectively.

\subsection{Backhaul Model\label{subsec:Backhaul-Model}}

We consider that mmWave is used to convey wireless backhaul for fixed-wing
UAVs owing to 1) higher frequency bands, which is non-overlapping
with those in UAV-GU channel, and 2) greater bandwidth and backhaul
capacity \cite{Ref_Background_mmWave_1,Ref_mmWave_model}. For instance,
it was tested in \cite{Ref_mmWave_parameter} that approximately 10Gbps
transmission rate with <20\% outage could be provided through 200m
mmWave transmissions over Ka band (28GHz), V band (60GHz) and E band
(73GHz) even if 3 blockages exist between mmWave transmitters and
receivers. Since the number of ground gateways is limited, we assume
that mmWave backhaul capacity is shared by the UAVs within the region
of radius $R_{\mathrm{m}}$ \cite{Ref_mmWave_model,backhaul_limit1,Ref_backhaul_model}.
Specifically, each UAV occupies equal proportion of backhaul resources.
According to \cite{backhaul_limit1,Ref_backhaul_model}, the backhaul
capacity for each UAV is given by
\begin{equation}
C_{\mathrm{b}}=\frac{G_{\mathrm{m}}\left(\vartheta\right)F_{\left|\varepsilon\right|}\left(\vartheta\right)C_{\mathrm{t}}}{\lambda_{\mathrm{a}}\pi R_{\mathrm{m}}^{2}},\:\mathrm{bits}/\left(\mathrm{s\cdot Hz}\right)\label{eq:backhaul model}
\end{equation}
where $C_{\mathrm{t}}$ denotes the normalized backhaul capacity in
$\mathrm{bits}/\left(\mathrm{s\cdot Hz}\right)$ and $\lambda_{\mathrm{a}}$
denotes the density of activated UAVs\footnote{Note that UAVs would keep inactivated with no backhaul requirement
if no GUs are connected.}. Therefore, the denominator $\lambda_{\mathrm{a}}\pi R_{\mathrm{m}}^{2}$
denotes the number of activated UAVs within the region.

In (\ref{eq:backhaul model}), $G_{\mathrm{m}}\left(\vartheta\right)=\frac{2\pi}{\vartheta}$
denotes the mmWave mainlobe gain and $F_{\left|\varepsilon\right|}\left(\vartheta\right)$
denotes the mmWave orientation effective function \cite{Ref_mmWave_model},
which is dependent on mmWave mainlobe beamwidth $\vartheta$ and orientation
error $\varepsilon$. Given $\left|\varepsilon\right|\leq\frac{\vartheta}{2}$,
backhaul could be effectively conveyed through mmWave. Otherwise,
UAV fails to obtain backhaul due to the orientation error. Supposing
that the mmWave mainlobe could cover the UAV hover region, we have
$\vartheta=2\arctan\left(\frac{R_{\mathrm{h}}}{h_{\mathrm{UAV}}}\right)$
under large $h_{\mathrm{UAV}}$. Accordingly, increasing the UAV hover
radius may alleviate the impact of orientation error, while reduces
the mmWave mainlobe gain.

\subsection{Performance metric\label{subsec:Performance-metric}}

In this work, we use coverage probability (CP) and network ST to evaluate
the performance of the UAV network. In particular, one GU is in coverage
when 1) the GU is in the projection area of UAVs and 2) the data transmission
from the associated UAV to the GU is successful. Without loss of generality,
we denote the link, which consists of $\mathrm{UAV}_{0}$ and $\mathrm{GU}_{0}$,
as the typical link. Accordingly, the CP of the typical link is defined
by
\begin{align}
\mathsf{CP} & =\mathbb{P}\left\{ \left\Vert \mathrm{UAV}_{0}-\mathrm{GU}_{0}\right\Vert <R_{\mathrm{p}},\mathsf{SIR}>\tau\right\} \nonumber \\
 & =p_{\mathrm{p}}\mathbb{P}\left\{ \mathsf{SIR}>\tau\left|E_{\mathrm{p}}\right.\right\} .\label{eq:define CP}
\end{align}
In (\ref{eq:define CP}), $E_{\mathrm{p}}$ denotes the event that
$\mathrm{GU}_{0}$ is within the projection area of $\mathrm{UAV}_{0}$
and we denote the corresponding projection probability as $p_{\mathrm{p}}=\mathbb{P}\left\{ \left\Vert \mathrm{UAV}_{0}-\mathrm{GU}_{0}\right\Vert <R_{\mathrm{p}}\right\} $.
If $E_{\mathrm{p}}$ occurs, $\mathbb{P}\left\{ \mathsf{SIR}>\tau\left|E_{\mathrm{p}}\right.\right\} $
denotes the transmission success probability, where $\mathsf{SIR}$
denotes the signal-to-interference ratio (SIR) at $\mathrm{GU}_{0}$
and $\tau$ is the SIR threshold.

Based on (\ref{eq:define CP}), network ST is defined by
\begin{align}
\mathsf{ST} & =\lambda_{\mathrm{a}}\mathsf{CP}\min\left\{ \log_{2}\left(1+\tau\right),C_{\mathrm{b}}\right\} \:\left[\mathrm{bits/(s\cdot Hz\cdot m^{2})}\right].\label{eq:define ST}
\end{align}
With the constraint of $C_{\mathrm{b}}$, $\mathsf{ST}$ captures
how many bits could be successfully conveyed over unit time, frequency
and area by the UAV network under limited backhaul.

\section{User Association in UAV Network\label{sec:Analysis}}

In this section, we evaluate the performance of the UAV network under
two user association rules, i.e., Semi-RTNA and RTNA. To capture the
impact of hover radius, we consider that the half-beamwidth $\Phi$
in (\ref{eq:antenna gain}) equals $\frac{\pi}{2}$. In consequence,
the radius of the projection area of each UAV approaches infinity,
i.e., $R_{\mathrm{p}}\rightarrow\infty$, and UAV antenna gain degenerates
into 1. Besides, GUs are always in the projection area of the UAVs,
i.e., $p_{\mathrm{p}}=1$ in (\ref{eq:define CP}).

Following the definitions of CP and network ST in Section \ref{subsec:Performance-metric},
the key to the analysis is to calculate the distribution of SIR at
$\mathrm{GU}_{0}$. However, the SIR distributions under Semi-RTNA
and RTNA rules are different, which will be discussed in the following.

\subsection{Semi-RTNA Rule}

With Semi-RTNA, the SIR at $\mathrm{GU}_{0}$ can be expressed as
\begin{align}
\mathsf{SIR}_{\mathrm{GU}_{0}}^{\mathrm{SR}}= & \frac{PH_{\mathrm{UAV}_{0}}d_{0}^{-\alpha}}{I^{\mathrm{SR}}},\label{eq:SIR semi-RTNA}
\end{align}
where $d_{0}$ denotes the distance from $\mathrm{UAV}_{0}$ to $\mathrm{GU}_{0}$
and $I^{\mathrm{SR}}=\underset{\tiny{\mathrm{UAV}_{i}\in\mathrm{\tilde{\Pi}_{UAV}}\backslash\mathrm{UAV}_{0}}}{\sum}PH_{\mathrm{UAV}_{i}}d_{i}^{-\alpha}$
denotes the interference stemming from other activated UAV cells.
$H_{\mathrm{UAV}_{i}}$ denotes the channel power gain due to small-scale
fading and $\mathrm{\tilde{\Pi}_{UAV}}$ denotes the set of activated
UAVs with density $\lambda_{\mathrm{a}}=q_{\mathrm{a}}\lambda$. Due
to UAV association, the UAV activation probability is given by \cite{Ref_active_probability}
\begin{align}
q_{\mathrm{a}} & =1-\left(1+\left(\mu\eta\right)^{-1}\right)^{-\mu},\label{eq:activation probability}
\end{align}
where $\eta=\lambda/\lambda_{\mathrm{GU}}$ denotes the ratio of UAV
density to GU density and $\mu$=3.5. Moreover, we suppose that the
mmWave orientation error $\varepsilon$ in (\ref{eq:backhaul model})
follows exponential distribution\cite{Ref_mmWave_model,Ref_mmWave_model_2}.
Therefore, the mmWave effective function is given by the following
lemma\footnote{Note that the results in Lemma \ref{lemma: error function} can be
feasibly extended to the cases, where other distributions are applied
to model the orientation error.}.

\begin{lemma}

Supposing that the absolute mmWave beam orientation error $\left|\varepsilon\right|$
follows exponential distribution truncated to $\left[0,\pi\right]$,
the mmWave effective function $F_{\left|\varepsilon\right|}\left(\vartheta\right)$
is given by
\begin{align}
F_{\left|\varepsilon\right|}\left(\vartheta\right) & =\frac{1-\mathrm{exp}\left(-\frac{\vartheta}{2\bar{\varepsilon}}\right)}{1-\mathrm{exp}\left(-\frac{\pi}{\bar{\varepsilon}}\right)},\:\vartheta\in\left(0,\frac{\pi}{2}\right)\label{eq:mmWave error function}
\end{align}
where $\bar{\varepsilon}$ denotes the mean of $\varepsilon$ (prior
to the truncation).

\label{lemma: error function}

\end{lemma}

\textit{Proof}: Please refer to Appendix \ref{subsec:Proof error function}.\qed

On average, the mmWave effective function $F_{\left|\varepsilon\right|}\left(\vartheta\right)$
in (\ref{eq:mmWave error function}) could capture the impact of mean
orientation error $\bar{\varepsilon}$ and mmWave mainlobe beamwidth
$\vartheta$ on the performance of mmWave transmission. Specifically,
$F_{\left|\varepsilon\right|}\left(\vartheta\right)$ is shown to
be reduced by either the increase of $\bar{\varepsilon}$ or the decrease
of $\vartheta$, which complies with intuition. According to Lemma
\ref{lemma: error function}, it can be shown that $G_{\mathrm{m}}\left(\vartheta\right)F_{\left|\varepsilon\right|}\left(\vartheta\right)$
in (\ref{eq:backhaul model}) is a decreasing function of the mmWave
mainlobe beamwidth $\vartheta$. Since $\vartheta=2\arctan\left(\frac{R_{\mathrm{h}}}{h_{\mathrm{UAV}}}\right)$,
increasing the UAV hover radius would potentially degrade the mmWave
backhaul capacity.

Aided by (\ref{eq:SIR semi-RTNA}), (\ref{eq:activation probability})
and Lemma \ref{lemma: error function}, CP and network ST under Semi-RTNA
could be obtained in the following proposition.

\begin{proposition}

Each UAV equipped with $\Phi=\frac{\pi}{2}$ directional antenna,
network ST of the fixed-wing UAV network under Semi-RTNA rule is given
by
\begin{align}
\mathsf{ST}_{\pi/2}^{\mathrm{SR}} & =\lambda_{\mathrm{a}}\mathsf{CP}_{\pi/2}^{\mathrm{SR}}\min\left\{ \log_{2}\left(1+\tau\right),C_{\mathrm{b}}\right\} .\label{eq:ST semi-RTNA omni}
\end{align}
In (\ref{eq:ST semi-RTNA omni}), $\mathsf{CP}_{\pi/2}^{\mathrm{SR}}$
is given by
\begin{align}
\mathsf{CP}_{\pi/2}^{\mathrm{SR}} & =\mathbb{E}_{r_{0},\theta,\Delta h}\left[\exp\left(-2\pi\lambda_{\mathrm{a}}\frac{\tau d_{0}^{\alpha}l^{2-\alpha}\omega_{1}\left(\alpha,\tau d_{0}^{\alpha}l^{-\alpha}\right)}{\alpha-2}\right)\right],\label{eq:CP semi-RTNA omni}
\end{align}
where $\theta$ denotes the hover angle, $d_{0}=\sqrt{r_{0}^{2}+R_{\mathrm{h}}^{2}-2r_{0}R_{\mathrm{h}}\cos\theta+\Delta h^{2}}$
and 
\[
l=\begin{cases}
\Delta h, & r_{0}\leq R_{\mathrm{h}}\\
\sqrt{\left(r_{0}-R_{\mathrm{h}}\right)^{2}+\Delta h^{2}}. & r_{0}\geq R_{\mathrm{h}}
\end{cases}
\]
If denoting $_{2}F_{1}\left(\cdotp,\cdotp;\cdotp;\cdotp\right)$ as
the standard Gaussian hypergeometric function, we have $\omega_{1}\left(x,y\right)={}_{2}F_{1}\left(1,1-\frac{2}{x};2-\frac{2}{x};-y\right)$.

The probability density functions (PDFs) of $r_{0}$, $\theta$ and
$\Delta h$ are, respectively, given by
\begin{align}
f_{r_{0}}\left(x\right) & =2\pi\lambda x\exp\left(-\pi\lambda x^{2}\right),\:x\in\left(0,\infty\right)\label{eq:PDF link length}
\end{align}
\begin{align}
f_{\theta}\left(x\right)= & \frac{1}{2\pi},\:x\in\left[0,2\pi\right]\label{eq:PDF hover angle}
\end{align}
\begin{align}
f_{\Delta h}\left(x\right) & =\frac{1}{\tilde{h}_{\mathrm{U}}-\tilde{h}_{\mathrm{L}}}.\:x\in\left[\tilde{h}_{\mathrm{L}},\tilde{h}_{\mathrm{U}}\right]\label{eq:PDF flying altitude}
\end{align}

\label{proposition: CP and ST Semi-RTNA omni}

\end{proposition}

\textit{Proof}: Please refer to Appendix \ref{subsec:Proof CP ST semi-RTNA omni}.\qed

It is observed from Proposition \ref{proposition: CP and ST Semi-RTNA omni}
that the CP and ST of UAV APs are dependent on the flight altitude
of UAVs, UAV hover radius and mmWave backhaul constraint, etc. Moreover,
it is shown from (\ref{eq:CP semi-RTNA omni}) that the expression
of CP is in complicated form, which is due to the varying flight altitude
of UAVs. To shed light on the impact of hover radius on the performance
of UAV network, we analyze the upper bound of CP given a fixed flight
altitude in the following corollary.

\begin{corollary}

Given a fixed UAV flight altitude, the conditional CP of $\mathrm{GU}_{0}$
is upper bounded by
\begin{align}
\hat{\mathsf{CP}}_{\pi/2\left|\Delta h\right.}^{\mathrm{SR}}= & \frac{\exp\left(-\pi\lambda\delta\left(\alpha,\tau,q_{\mathrm{a}}\right)\left(R_{\mathrm{h}}^{2}+\Delta h^{2}\right)\right)}{1+\delta\left(\alpha,\tau,q_{\mathrm{a}}\right)}\nonumber \\
\times & \left(1+\frac{\delta\left(\alpha,\tau,q_{\mathrm{a}}\right)\pi\sqrt{\lambda}R_{\mathrm{h}}}{\left(1+\delta\left(\alpha,\tau,q_{\mathrm{a}}\right)\right)^{\frac{3}{2}}}\left(1+\mathrm{Erf}\left(\frac{\delta\left(\alpha,\tau,q_{\mathrm{a}}\right)\sqrt{\pi\lambda}R_{\mathrm{h}}}{\sqrt{1+\delta\left(\alpha,\tau,q_{\mathrm{a}}\right)}}\right)\right)\right),\label{eq:CP semi-RTNA omni UB}
\end{align}
where $\delta\left(\alpha,\tau,q_{\mathrm{a}}\right)=\frac{2q_{\mathrm{a}}\tau\omega_{1}\left(\alpha,\tau\right)}{\alpha-2}$
and $\mathrm{Erf}\left(\cdot\right)$ denotes the standard error function.

\label{corollary: CP and ST Semi-RTNA omni upper bound}

\end{corollary}

\textit{Proof}: Please refer to Appendix \ref{subsec:Proof CP ST semi-RTNA omni UB}.\qed

Following Corollary \ref{corollary: CP and ST Semi-RTNA omni upper bound},
it can be shown that $\frac{\partial\hat{\mathsf{CP}}_{\pi/2\left|\Delta h\right.}^{\mathrm{SR}}}{\partial R_{\mathrm{h}}}<0$.
This indicates that increasing the UAV hover radius would deteriorate
the performance of UAV network under Semi-RTNA rule\footnote{Note that deconditioning $\Delta h$ through calculating the expectation
of $\Delta h$ will not disprove the above results.}. On the one hand, when the associated UAV is departing from the GU,
the transmission distance will be increased, thereby reducing the
desired signal power. On the other hand, when the neighboring UAVs
are approaching the intended GU, inter-cell interference becomes severe
and dominates the performance of air-to-ground transmission.

\begin{figure}[t]
\centering{}\subfloat[CP v.s. $R_{\mathrm{h}}$.]{\begin{centering}
\includegraphics[width=3in]{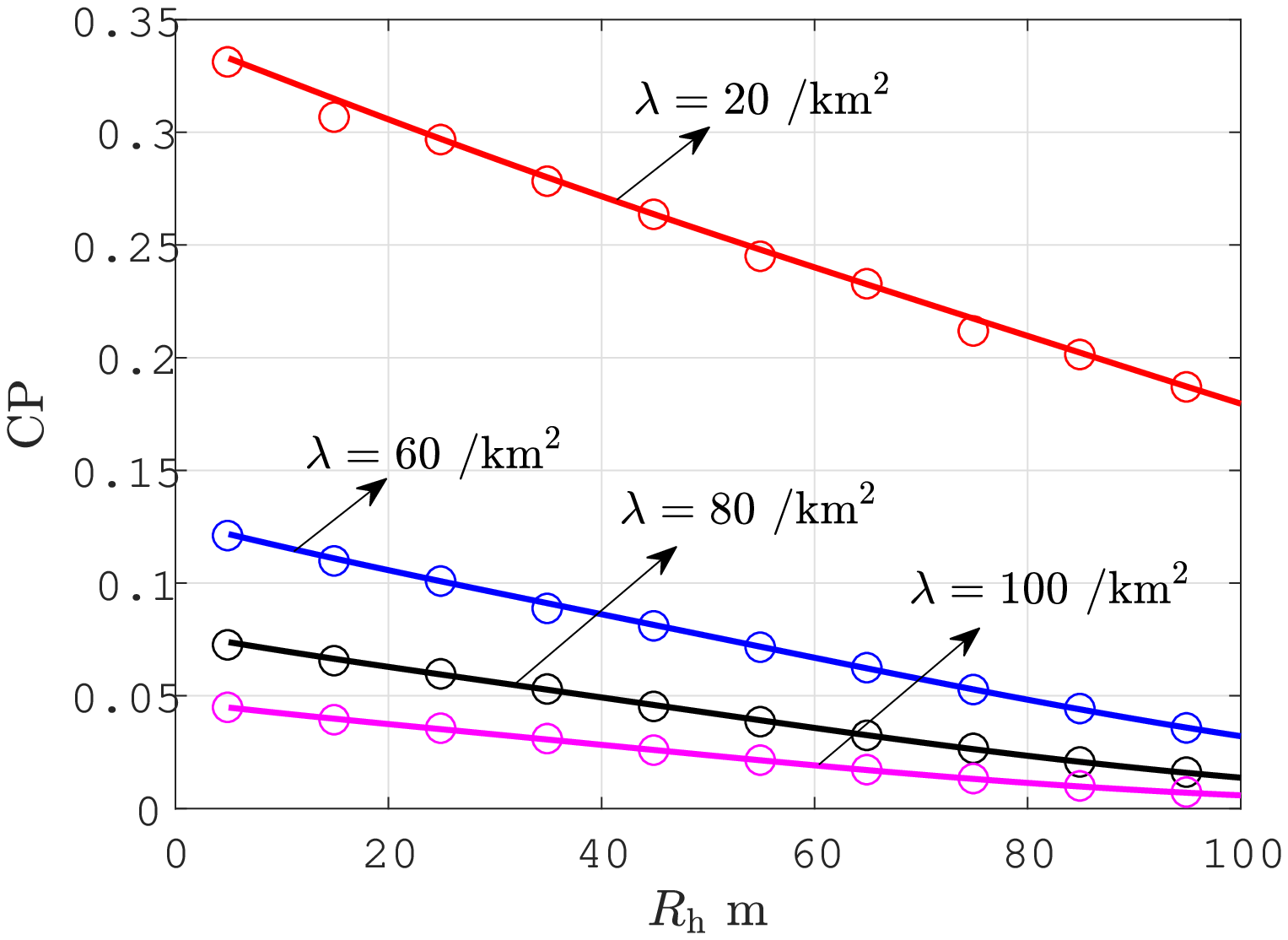}
\par\end{centering}
}\subfloat[ST v.s. $R_{\mathrm{h}}$.]{\begin{centering}
\includegraphics[width=3in]{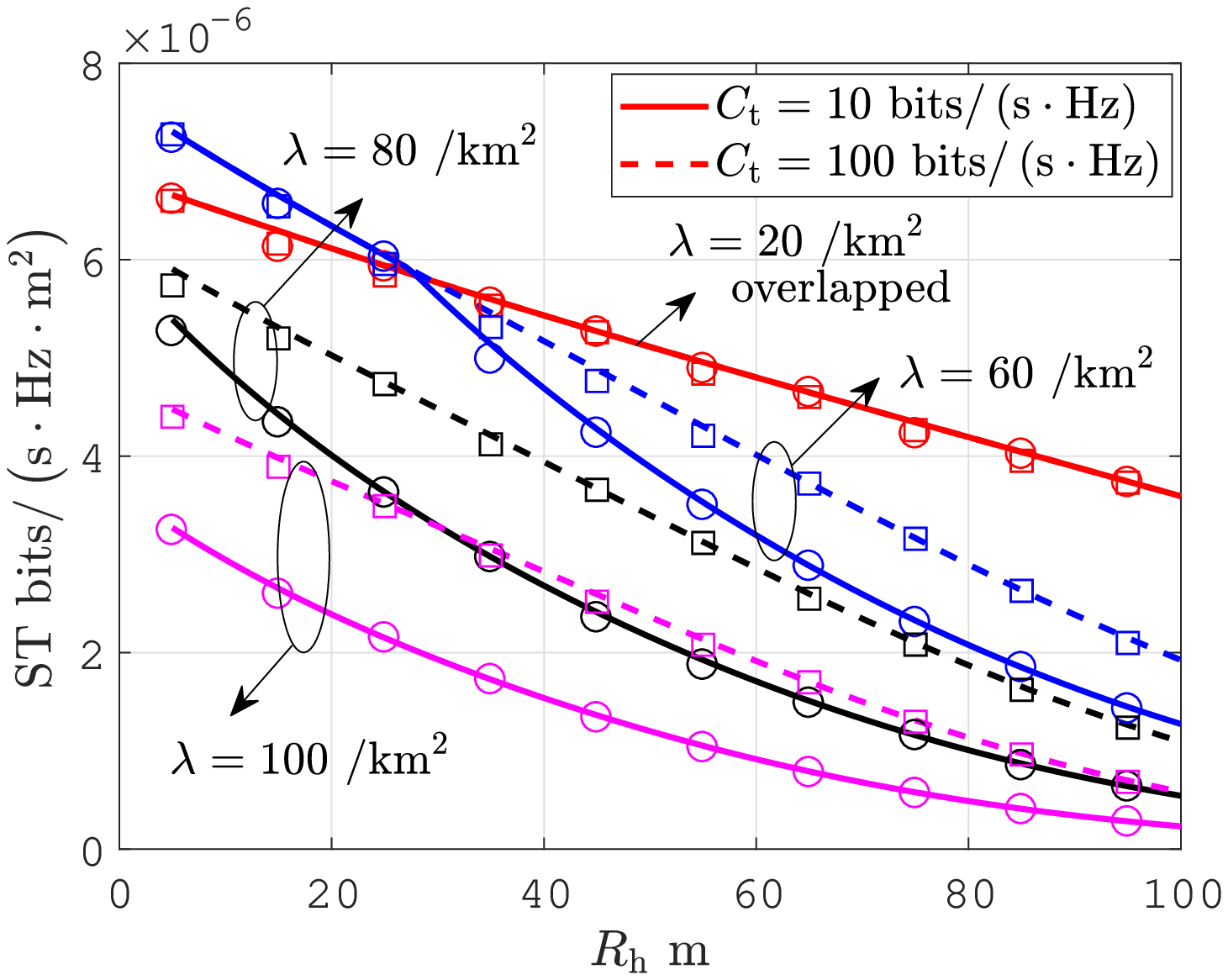}
\par\end{centering}
}\caption{\label{fig:Impact of hover radius}CP and network ST under different
hover radius $R_{\mathrm{h}}$. For system parameters, we set $P=30$dBm,
$\Delta h\sim\mathcal{U}\left(90\mathrm{m},110\mathrm{m}\right)$,
$\tau=1$, $\alpha=3.5$ and $\lambda_{\mathrm{GU}}=1\times10^{4}$
user/$\mathrm{km}^{2}$. For backhaul parameters, we set $R_{\mathrm{m}}=500$m.
In this and following figures, numerical and simulation results are
denoted by lines and markers, respectively.}
\end{figure}

To verify the above results, we plot Fig. \ref{fig:Impact of hover radius},
which shows CP and network ST as a function of UAV hover radius $R_{\mathrm{h}}$
under different UAV deployment densities $\lambda$ and backhaul capacity
constraints $C_{\mathrm{t}}$. It can be seen that either CP or network
ST decreases with $R_{\mathrm{h}}$ under different UAV densities.
In particular, network ST is more significantly degraded given smaller
$C_{\mathrm{t}}$, e.g., $C_{\mathrm{t}}=10$bit/$\left(\mathrm{s}\cdot\mathrm{Hz}\right)$,
and greater $\lambda$, e.g., $\lambda>60/\mathrm{km}^{2}$. As discussed,
the reason is that the hovering feature of UAVs would result in the
connection discontinuity of GUs. In consequence, the UAV-GU transmission
is more vulnerable to air-to-ground interference especially when more
UAVs are deployed. Moreover, the mmWave mainlobe beamwidth has to
be adaptively increased with the UAV hovering range so as to provide
sufficient backhaul, which nevertheless reduces the mmWave mainlobe
gain and limits the backhaul link capacity.

\subsection{RTNA Rule}

If frequent handover is allowed under RTNA, the performance of UAV
network could intuitively be improved since GUs always connect to
the closest hovering UAVs. In this light, we analyze CP and network
ST under RTNA in the following proposition.

\begin{proposition}

Each UAV equipped with $\Phi=\frac{\pi}{2}$ directional antenna,
network ST of the fixed-wing UAV network under RTNA rule is given
by
\begin{align}
\mathsf{ST}_{\pi/2}^{\mathrm{R}} & =\lambda_{\mathrm{a}}\mathsf{CP}_{\pi/2}^{\mathrm{R}}\min\left\{ \log_{2}\left(1+\tau\right),C_{\mathrm{b}}\right\} .\label{eq:ST RTNA omni}
\end{align}
In (\ref{eq:ST RTNA omni}), $\mathsf{CP}_{\pi/2}^{\mathrm{R}}$ is
given by
\begin{align}
\mathsf{CP}_{\pi/2}^{\mathrm{R}} & =\frac{\mathrm{Erf}\left(\sqrt{\delta\left(\alpha,\tau,q_{\mathrm{a}}\right)}\tilde{h}_{\mathrm{U}}\right)-\mathrm{Erf}\left(\sqrt{\delta\left(\alpha,\tau,q_{\mathrm{a}}\right)}\tilde{h}_{\mathrm{L}}\right)}{2\sqrt{\lambda\delta\left(\alpha,\tau,q_{\mathrm{a}}\right)}\left(1+\delta\left(\alpha,\tau,q_{\mathrm{a}}\right)\right)\left(\tilde{h}_{\mathrm{U}}-\tilde{h}_{\mathrm{L}}\right)}.\label{eq:CP RTNA omni}
\end{align}

\label{proposition: CP and ST RTNA omni}

\end{proposition}

\textit{Proof}: Please refer to Appendix \ref{subsec:Proof CP ST RTNA omni}.\qed

It is shown in Proposition \ref{proposition: CP and ST RTNA omni}
that the CP under RTNA is independent of the UAV hover radius. This
is because, when hover radius is increased or decreased, GUs could
always be associated with the closest UAV. In consequence, the problem
raised by the hover of UAVs in Semi-RTNA rule will be alleviated,
i.e., dominant interference due to approaching neighboring UAVs and
reduced desired signal power due to departing associated UAVs.

Even though the results in Proposition \ref{proposition: CP and ST RTNA omni}
are still in complicated forms, we make a comparison of the two rules
with the aid of Corollary \ref{corollary: CP and ST Semi-RTNA omni upper bound}.
Deconditioning $\Delta h$ in (\ref{eq:CP semi-RTNA omni UB}), we
have
\begin{align}
\mathbb{E}_{\Delta h}\left[\hat{\mathsf{CP}}_{\pi/2\left|\Delta h\right.}^{\mathrm{SR}}\right] & \overset{\left(\mathrm{a}\right)}{<}\mathbb{E}_{\Delta h}\left[\frac{\exp\left(-\pi\lambda\delta\left(\alpha,\tau,q_{\mathrm{a}}\right)\Delta h^{2}\right)}{1+\delta\left(\alpha,\tau,q_{\mathrm{a}}\right)}\right]\nonumber \\
 & =\mathsf{CP}_{\pi/2}^{\mathrm{R}},\label{eq:comparison RTNA semi-RTNA}
\end{align}
where (a) follows by setting $R_{\mathrm{h}}=0$ because $\hat{\mathsf{CP}}_{\pi/2\left|\Delta h\right.}^{\mathrm{SR}}$
is a decreasing function of $R_{\mathrm{h}}$. Since $\mathsf{CP}_{\pi/2}^{\mathrm{SR}}<\mathbb{E}_{\Delta h}\left[\hat{\mathsf{CP}}_{\pi/2\left|\Delta h\right.}^{\mathrm{SR}}\right]$
(see Corollary \ref{corollary: CP and ST Semi-RTNA omni upper bound})
and the impact of mmWave backhaul is identical to RTNA and Semi-RTNA,
(\ref{eq:comparison RTNA semi-RTNA}) indicates that RTNA outperforms
Semi-RTNA in terms of CP and network ST.

\begin{figure}[t]
\begin{centering}
\subfloat[\label{fig:Impact of hover radius - CP}CP v.s. $\lambda$.]{\begin{centering}
\includegraphics[width=3in]{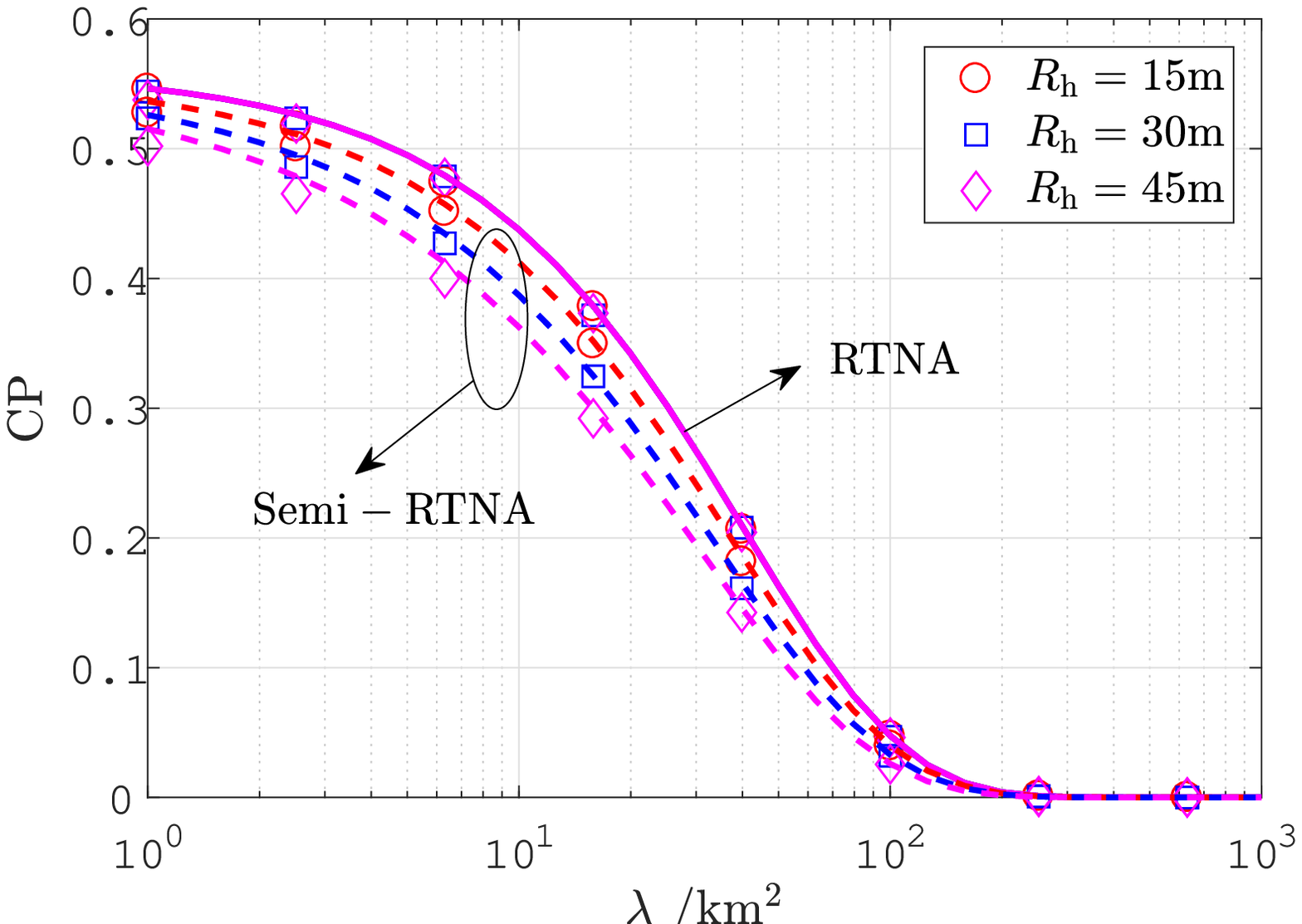}
\par\end{centering}
}\subfloat[\label{fig:Impact of hover radius - ST}ST v.s. $\lambda$.]{\begin{centering}
\includegraphics[width=3in]{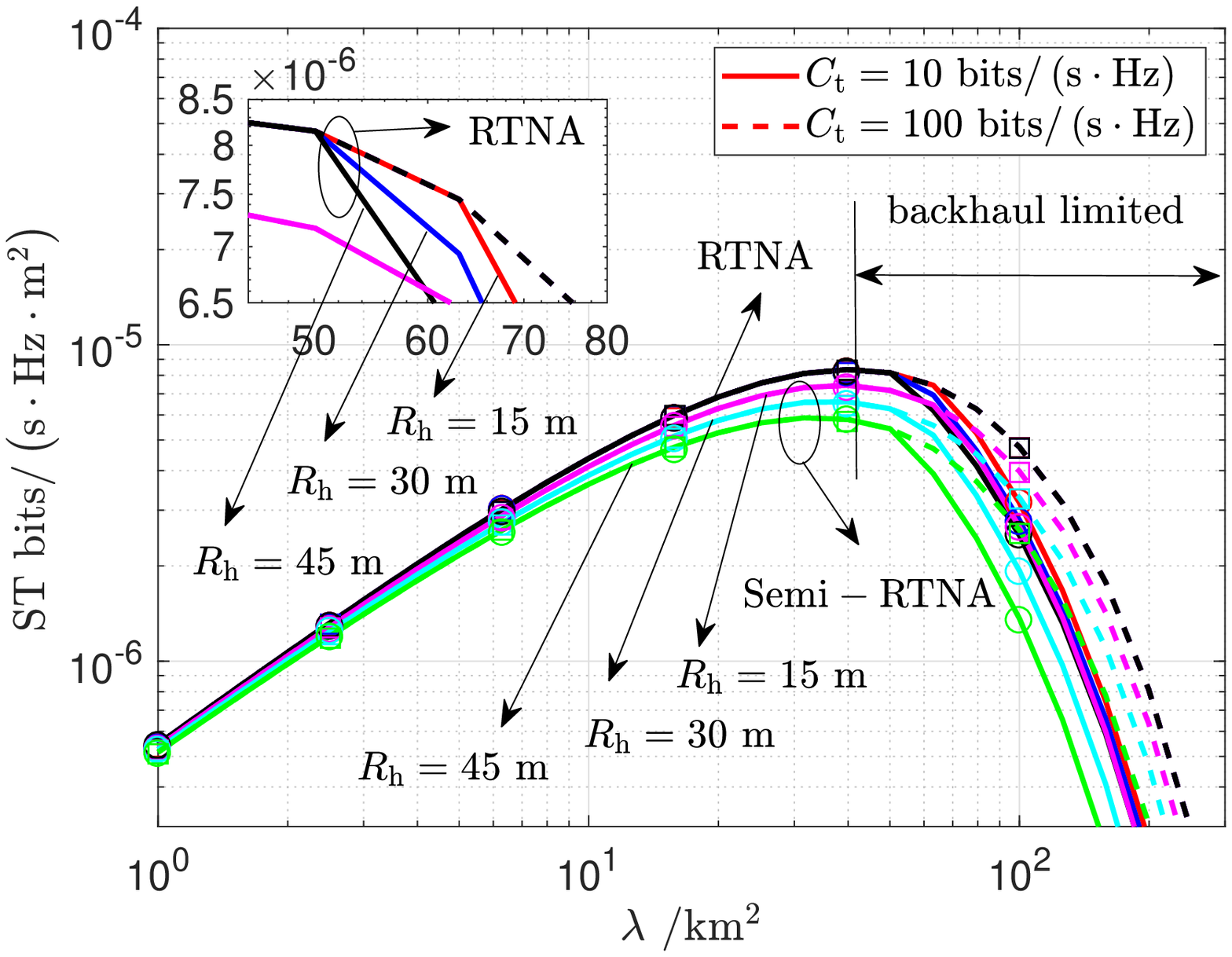}
\par\end{centering}
}
\par\end{centering}
\caption{\label{fig:comparison of semi-RTNA and RTNA}CP and Network ST under
different UAV density $\lambda$. Except for $R_{\mathrm{h}}$, system
parameters are identically set as those in Fig. \ref{fig:Impact of hover radius}.}
\end{figure}

In Fig. \ref{fig:comparison of semi-RTNA and RTNA}, we plot CP and
network ST with varying UAV deployment density $\lambda$ under Semi-RTNA
and RTNA. It is shown that RTNA outperforms Semi-RTNA in terms of
CP and network ST. As indicated by Proposition \ref{proposition: CP and ST RTNA omni},
the CP is shown to be independent of the UAV hover radius under RTNA
in Fig. \ref{fig:Impact of hover radius - CP}. Moreover, when $\lambda$
is small, the network ST obtained by RTNA overlaps under different
hover radii, which is shown in Fig. \ref{fig:Impact of hover radius - ST}.
When $\lambda$ increases, limited wireless backhaul will be shared
by more UAVs. In consequence, it can be seen in Fig. \ref{fig:Impact of hover radius - ST}
that network ST is more significantly degraded (solid lines) in the
backhaul-limited region. Moreover, it can be seen that a greater hover
radius would lead to a more rapid decrease of network ST. As discussed
earlier, the reason is that the mmWave link capacity will be decreased
with the high mobility of UAVs due to reduced mainlobe gain. Consequently,
the UAV network performance is more likely to be dominated by limited
backhaul. 

In addition, compared to the terrestrial network, in which the coverage
of terrestrial APs is limited, the coverage of UAVs is greatly enhanced
due to higher deployment altitude, which increases the overlapping
area of the UAV cells. As a result, it is shown in Fig. \ref{fig:comparison of semi-RTNA and RTNA}
that network ST begins to diminish with UAV density $\lambda$ even
when $\lambda$ is small, e.g., $\lambda>40/\mathrm{km}^{2}$, due
to the inter-cell interference caused by overlapping UAV cells. This
can be analytically verified through Proposition \ref{proposition: CP and ST RTNA omni}.
For this reason, it is crucial to investigate how to alleviate the
overwhelming interference in the UAV network, which is discussed in
Section \ref{sec:Optimization}.

\section{Analysis and Optimization of Directional-Antenna UAV Network\label{sec:Optimization}}

The adjustment of directional-antenna UAV beamwidth is of great potential
to avoid the overlap of UAV cells and mitigate the inter-cell interference.
In this light, we first evaluate the performance of the directional-antenna
UAV network under the two user association rules in the following.

\subsection{Performance Analysis}

When Semi-RTNA is applied, the SIR at $\mathrm{GU}_{0}$ can be expressed
as
\begin{align}
\mathsf{\hat{SIR}}_{\mathrm{GU}_{0}}^{\mathrm{SR}}= & \frac{PH_{\mathrm{UAV}_{0}}G\left(\phi,\varphi\right)d_{0}^{-\alpha}}{\hat{I}^{\mathrm{SR}}},\label{eq:SIR semi-RTNA dir}
\end{align}
where $G\left(\phi,\varphi\right)$ is given by (\ref{eq:antenna gain})
in Section \ref{subsec:Channel Model}, $\hat{I}^{\mathrm{SR}}=\underset{\tiny{\mathrm{UAV}_{i}\in\mathrm{\tilde{\Pi}_{UAV}}\backslash\mathrm{UAV}_{0}}}{\sum}PH_{\mathrm{UAV}_{i}}G\left(\phi,\varphi\right)d_{i}^{-\alpha}$
denotes the interference stemming from the overlapping activated UAV
cells. According to (\ref{eq:define CP}), in addition to transmission
success, CP is as well dependent on whether GUs are within the projection
area of the associated UAVs. Since $\left\Vert \mathrm{UAV}_{0}-\mathrm{GU}_{0}\right\Vert =\sqrt{r_{0}^{2}+R_{\mathrm{h}}^{2}-2r_{0}R_{\mathrm{h}}\cos\theta}$
is the 2D distance from $\mathrm{UAV}_{0}$ to $\mathrm{GU}_{0}$,
the projection probability $p_{\mathrm{p}}$ is given by (\ref{eq:projection probability})
to facilitate the analysis.
\begin{align}
p_{\mathrm{p}} & =\mathbb{P}\left(\left\Vert \mathrm{UAV}_{0}-\mathrm{GU}_{0}\right\Vert <R_{\mathrm{p}}\right)\nonumber \\
 & =\underset{\left\Vert \mathrm{UAV}_{0}-\mathrm{GU}_{0}\right\Vert <R_{\mathrm{p}}}{\iint}r_{0}\lambda\exp\left(-\pi\lambda r_{0}^{2}\right)d\theta dr_{0}.\label{eq:projection probability}
\end{align}

Following (\ref{eq:SIR semi-RTNA dir}) and (\ref{eq:projection probability}),
CP and network ST are given in Proposition \ref{proposition: CP and ST Semi-RTNA dir}.

\begin{proposition}

Each UAV equipped with directional-antenna $\Phi\in\left(0,\frac{\pi}{2}\right]$,
network ST of the fixed-wing UAV network under Semi-RTNA rule is given
by
\begin{align}
\mathsf{ST}_{\mathrm{dir}}^{\mathrm{SR}} & =\lambda_{\mathrm{a}}\mathsf{CP}_{\mathrm{dir}}^{\mathrm{SR}}\min\left\{ \log_{2}\left(1+\tau\right),C_{\mathrm{b}}\right\} .\label{eq:ST semi-RTNA dir}
\end{align}
In (\ref{eq:ST semi-RTNA dir}), $\mathsf{CP}_{\mathrm{dir}}^{\mathrm{SR}}$
is given by
\begin{align}
\mathsf{CP}_{\mathrm{dir}}^{\mathrm{SR}} & =\mathbb{E}_{r_{0},\theta,\Delta h}\left[p_{\mathrm{p}}\mathsf{CP}_{\mathrm{dir}\left|r_{0},\theta,\Delta h\right.}^{\mathrm{SR}}\left|E_{\mathrm{p}}\right.\right].\label{eq:CP semi-RTNA dir}
\end{align}
In (\ref{eq:CP semi-RTNA dir}), $\mathsf{CP}_{\mathrm{dir}\left|r_{0},\theta,\Delta h\right.}^{\mathrm{SR}}$
is given by
\begin{align}
\mathsf{CP}_{\mathrm{dir}\left|r_{0},\theta,\Delta h\right.}^{\mathrm{SR}}= & \begin{cases}
\exp\left[-\pi p_{\mathrm{p}}\lambda_{\mathrm{a}}\left(\hat{R}_{\mathrm{p}}^{2}\omega_{2}\left(\alpha,\frac{\hat{R}_{\mathrm{p}}^{\alpha}}{\tau d_{0}^{\alpha}}\right)-\hat{d}_{0}^{2}\omega_{2}\left(\alpha,\frac{\hat{d}_{0}^{\alpha}}{\tau d_{0}^{\alpha}}\right)\right)\right], & r_{0}\geq R_{\mathrm{h}}\\
\exp\left[-\pi p_{\mathrm{p}}\lambda_{\mathrm{a}}\left(\hat{R}_{\mathrm{p}}^{2}\omega_{2}\left(\alpha,\frac{\hat{R}_{\mathrm{p}}^{\alpha}}{\tau d_{0}^{\alpha}}\right)-\Delta h^{2}\omega_{2}\left(\alpha,\frac{\Delta h^{\alpha}}{\tau d_{0}^{\alpha}}\right)\right)\right], & \mathrm{otherwise}
\end{cases}\label{eq:CP semi-RTNA dir ex}
\end{align}
where $\hat{R}_{\mathrm{p}}=\sqrt{R_{\mathrm{p}}^{2}+\Delta h^{2}}$,
$\omega_{2}\left(x,y\right)={}_{2}F_{1}\left(1,\frac{2}{x};1+\frac{2}{x};-y\right)$,
$\hat{d}_{0}=\sqrt{\left(r_{0}-R_{\mathrm{h}}\right)^{2}+\Delta h^{2}}$
and $d_{0}=\sqrt{r_{0}^{2}+R_{\mathrm{h}}^{2}-2r_{0}R_{\mathrm{h}}\cos\theta+\Delta h^{2}}$.

\label{proposition: CP and ST Semi-RTNA dir}

\end{proposition}

\textit{Proof}: Please refer to Appendix \ref{subsec:Proof CP ST semi-RTNA dir}.\qed

Based on the results in Proposition \ref{proposition: CP and ST Semi-RTNA dir},
the impact of UAV projection radius $R_{\mathrm{p}}$, which is dependent
on UAV flight altitude and beamwidth, on the performance of UAV network
under Semi-RTNA can be revealed. To better illustrate the impact,
we give Lemma \ref{lemma: increasing function} in the following.

\begin{lemma}

Define $\Psi\left(x\right)=x^{2}{}_{2}F_{1}\left(1,\frac{2}{z};1+\frac{2}{z};-bx^{z}\right)$
$\left(x\geq0\right)$, where $b\left(>0\right)$ and $z\left(>0\right)$
are constant. $\Psi\left(x\right)$ is an increasing function of $x$.

\label{lemma: increasing function}

\end{lemma}

\textit{Proof}: Please refer to Appendix \ref{subsec:Proof increasing function}.\qed

Following the projection probability in (\ref{eq:projection probability}),
more GUs can be covered by UAVs under a greater projection radius
$R_{\mathrm{p}}$ (or equivalently UAV beamwidth) by increasing the
projection probability. According to Lemma \ref{lemma: increasing function},
$\hat{R}_{\mathrm{p}}^{2}\omega_{2}\left(\alpha,\frac{\hat{R}_{\mathrm{p}}^{\alpha}}{\tau d_{0}^{\alpha}}\right)$
in (\ref{eq:CP semi-RTNA dir ex}) is an increasing function of $\hat{R}_{\mathrm{p}}=\sqrt{R_{\mathrm{p}}^{2}+\Delta h^{2}}$.
Therefore, it indicates from Proposition \ref{proposition: CP and ST Semi-RTNA dir}
that increasing $R_{\mathrm{p}}$ may as well reduce the coverage
probability by increasing $\hat{R}_{\mathrm{p}}$ in (\ref{eq:CP semi-RTNA dir ex}).
Intuitively, this is due to the degrading inter-cell interference
among different UAV cells. For this reason, there is a tradeoff on
adjusting the UAV projection radius in practical UAV network. 

Based on Proposition \ref{proposition: CP and ST Semi-RTNA dir},
we further analyze the network ST of the directional-antenna UAV network
under RTNA.

\begin{proposition}

Each UAV equipped with directional-antenna $\Phi\in\left(0,\frac{\pi}{2}\right]$,
network ST of the fixed-wing UAV network under RTNA rule is given
by
\begin{align}
\mathsf{ST}_{\mathrm{dir}}^{\mathrm{R}} & =\lambda_{\mathrm{a}}\mathsf{CP}_{\mathrm{dir}}^{\mathrm{R}}\min\left\{ \log_{2}\left(1+\tau\right),C_{\mathrm{b}}\right\} .\label{eq:ST RTNA dir}
\end{align}
In (\ref{eq:ST RTNA dir}), $\mathsf{CP}_{\mathrm{dir}}^{\mathrm{R}}$
is given by
\begin{align}
\mathsf{CP}_{\mathrm{dir}}^{\mathrm{R}} & =\mathbb{E}_{\Delta h,r_{0}<R_{\mathrm{p}}}\left[\hat{p}_{\mathrm{p}}\exp\left(-\pi\hat{p}_{\mathrm{p}}\lambda_{\mathrm{a}}\left(\hat{R}_{\mathrm{p}}^{2}\omega_{2}\left(\alpha,\frac{\hat{R}_{\mathrm{p}}^{\alpha}}{\tau\check{d}_{0}^{\alpha}}\right)-\check{d}_{0}^{2}\omega_{2}\left(\alpha,\frac{1}{\tau}\right)\right)\right)\right],\label{eq:CP RTNA dir}
\end{align}
where $\hat{p}_{\mathrm{p}}=1-\exp\left(\pi\lambda R_{\mathrm{p}}^{2}\right)$
and $\check{d}_{0}=\sqrt{r_{0}^{2}+\Delta h^{2}}$.

\label{proposition: CP and ST RTNA dir}

\end{proposition}

\textit{Proof}: Please refer to Appendix \ref{subsec:Proof CP ST RTNA dir}.\qed

Similarly as the results under Semi-RTNA in Proposition \ref{proposition: CP and ST Semi-RTNA dir},
there is also a tradeoff on setting the projection radius $R_{\mathrm{p}}$
under RTNA. According to (\ref{eq:CP RTNA dir}), increasing $R_{\mathrm{p}}$
may enhance the UAV coverage, while degrades the transmission success
probability by introducing more inter-cell interference. Based on
Propositions \ref{proposition: CP and ST Semi-RTNA dir} and \ref{proposition: CP and ST RTNA dir},
we make a comparison of Semi-RTNA and RTNA with directional-antenna
equipped by each UAV in the following.

\begin{figure*}[t]
\begin{centering}
\subfloat[\label{fig:Impact of beamwidth a}$\mathsf{ST}$ v.s. $\Phi$.]{\begin{centering}
\includegraphics[width=3in]{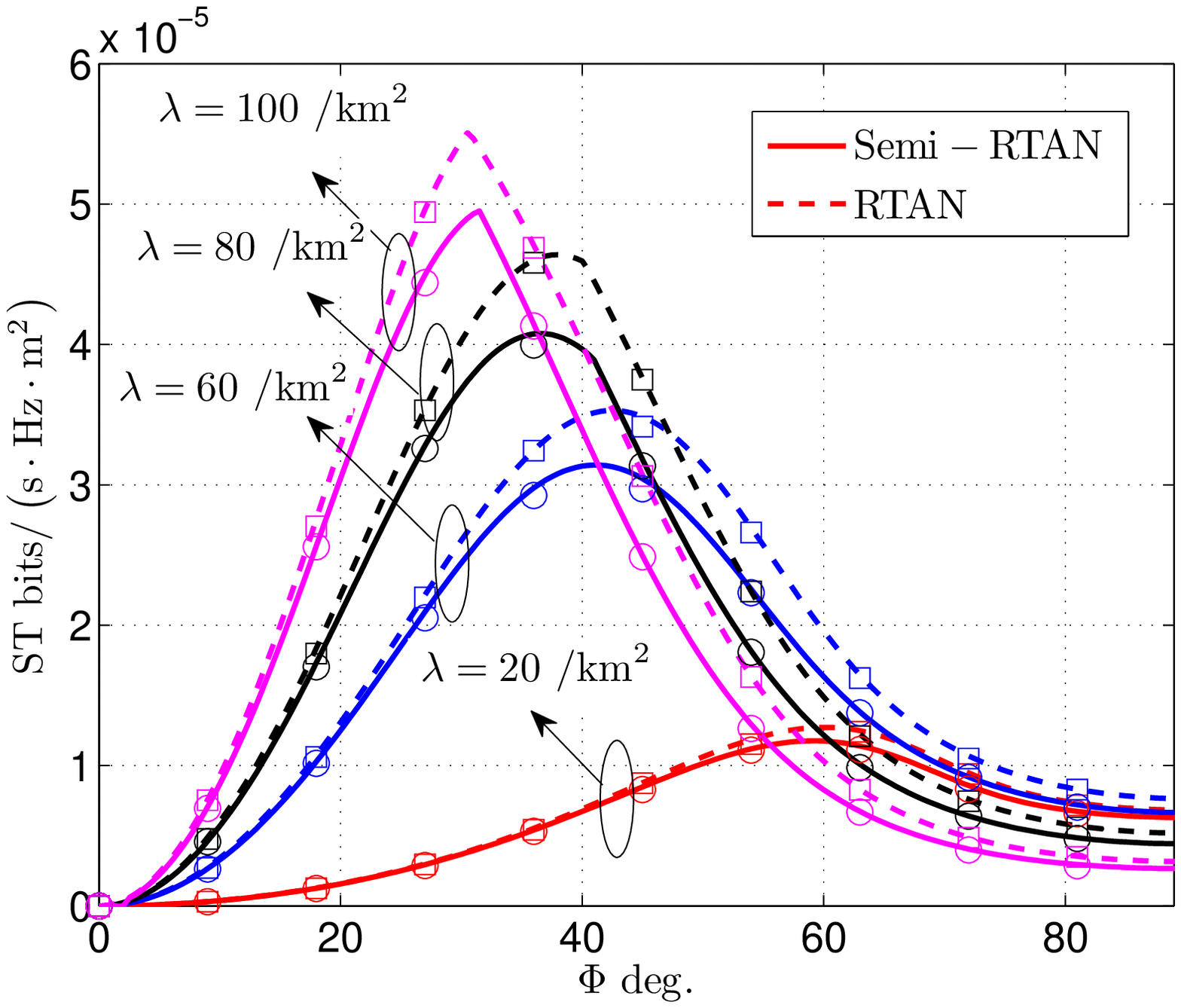}
\par\end{centering}
}\subfloat[\label{fig:Impact of beamwidth b}$\mathsf{ST}$ v.s. $\lambda$.]{\begin{centering}
\includegraphics[width=3in]{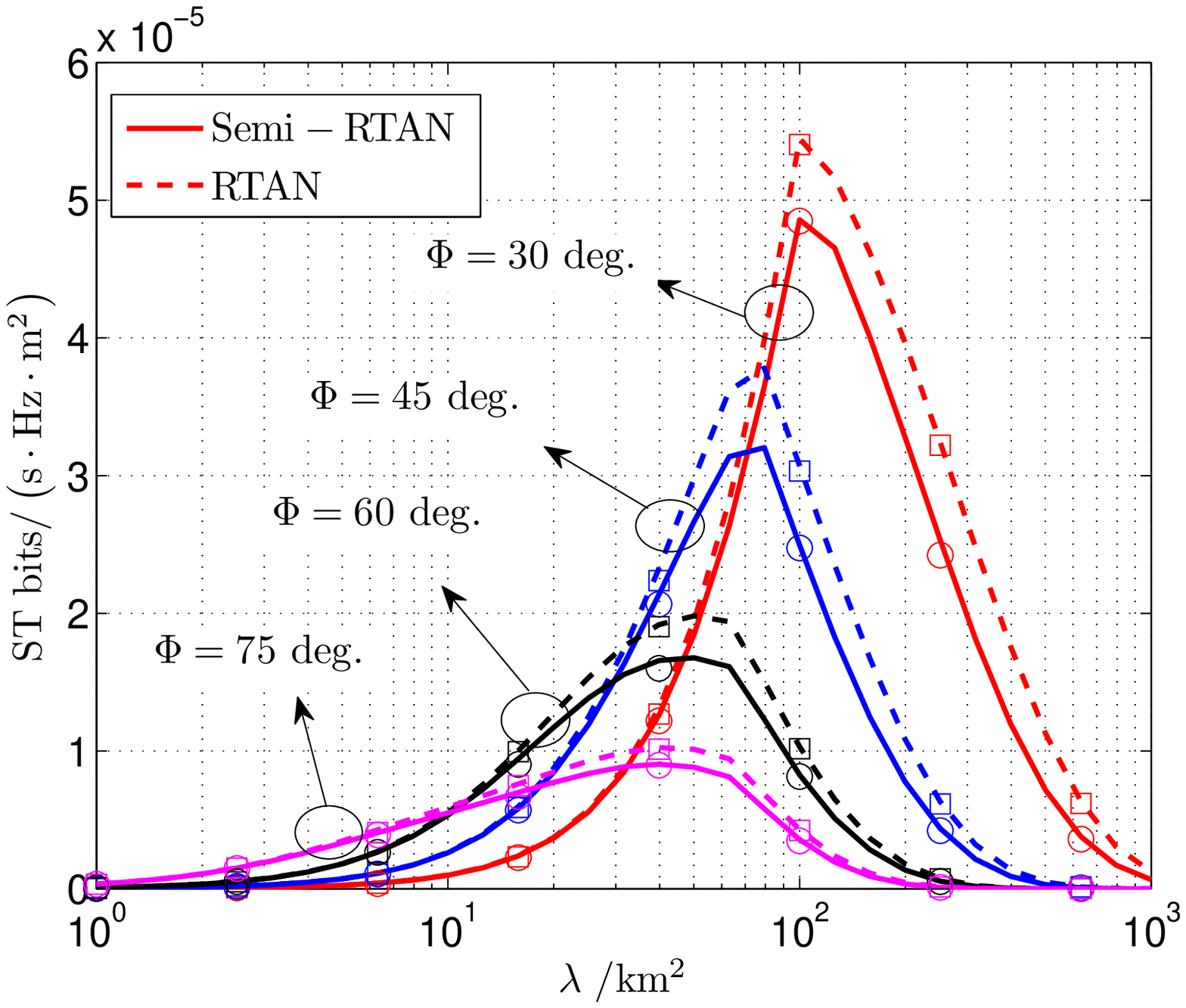}
\par\end{centering}
}
\par\end{centering}
\caption{\label{fig:Impact of beamwidth}Network ST under Semi-RTNA and RTNA.
System parameters are identically set as those in Fig. \ref{fig:Impact of hover radius}.
In addition, we set $C_{\mathrm{t}}=10$bit/$\left(\mathrm{s}\cdot\mathrm{Hz}\right)$.}
\end{figure*}

Fig. \ref{fig:Impact of beamwidth} plots network ST under Semi-RTNA
and RTNA. In particular, Fig. \ref{fig:Impact of beamwidth a} plots
network ST with varying UAV beamwidth under different UAV deployment
densities. It is observed that increasing the UAV beamwidth could
first enhance network ST since a greater air-to-ground coverage could
be provided. Nevertheless, network ST will be degraded due to the
dominant air-to-ground interference if UAV beamwidth further increases.
Meanwhile, it is shown in Figs. \ref{fig:Impact of beamwidth a} and
\ref{fig:Impact of beamwidth b} that a greater network ST can always
be obtained under RTNA. Moreover, the optimal UAV beamwidth, which
maximizes the network ST, inversely grows with the UAV deployment
density. This indicates that the optimal UAV beamwidth is critically
dependent on the UAV deployment density.

It is shown in Fig. \ref{fig:Impact of beamwidth} that appropriately
decreasing the UAV beamwidth could reduce the overlap of UAV projection
areas and therefore is of great potential in mitigating the inter-cell
interference and enhancing network ST. For this reason, we study how
to optimize the UAV beamwidth in the following.

\subsection{Performance Optimization}

In this part, we intend to maximize the network ST through tuning
the UAV beamwidth. To this end, we first formulate the optimization
problem as follows.

\begin{align}
\Phi^{*} & =\arg\underset{\Phi}{\max}\:\mathsf{ST}_{\mathrm{dir}}\left(\Phi\right)\label{eq:optimization problem}\\
\mathrm{s.t.} & \Phi\in\left(0,\frac{\pi}{2}\right].\nonumber 
\end{align}

From Propositions \ref{proposition: CP and ST Semi-RTNA dir} and
\ref{proposition: CP and ST RTNA dir}, it is shown that the UAV half-beamwidth
$\Phi$ has a complicated impact of network ST under either Semi-RTNA
or RTNA rule. As a consequence, it is difficult to obtain a closed-form
solution for the optimal half-beamwidth $\Phi^{*}$. Even if the optimization
problem in (\ref{eq:optimization problem}) is of single variable,
numerically solving it may consume much time in practical applications.
On this account, we propose a simple but effective PAE policy to tune
the UAV beamwidth in the following.

It is shown in Fig. \ref{fig:Impact of beamwidth a} that the optimal
UAV beamwidth, which maximizes network ST, always decreases with the
UAV deployment density. Moreover, it can be readily shown from (\ref{eq:PDF link length})
that the average 2D UAV-GU distance $\bar{r_{0}}$ follows $\bar{r_{0}}\propto\frac{1}{\sqrt{\lambda}}$.
Therefore, the concept of the PAE policy is that the UAV projection
radius $R_{\mathrm{p}}$ should be inversely proportional to the square
of activated UAV density $\lambda_{\mathrm{a}}$\footnote{Note that no coverage will be provided by the inactivated UAVs, which
are connected by no GUs.}. Especially, we have

\begin{align}
R_{\mathrm{p}}= & \frac{C}{\sqrt{\lambda_{\mathrm{a}}}},\label{eq:optimized projection radius}
\end{align}
where $C\left(>0\right)$ is defined as the \textit{projection scaling
parameter}. A greater $C$ will result in a greater projection radius
and equivalently UAV beamwidth. Therefore, supposing that the density
of UAV APs is small and inter-cell interference is moderate, increasing
$C$ would greatly increase the coverage of UAV APs and improve the
UAV network performance. Following (\ref{eq:optimized projection radius}),
the half-beamwidth of UAV is tuned as $\Phi^{*}=\arctan\left(\frac{C}{\Delta h\sqrt{\lambda_{\mathrm{a}}}}\right)$.
Note that the projection radius of each UAV could be identically set
as (\ref{eq:optimized projection radius}) through adjusting the beamwidth
even if UAVs are in different flight altitudes. Moreover, perfect
CSI is not required by the PAE policy. Therefore, the proposed policy
can be applied in practical fixed-wing UAV networks, where it is difficult
to acquire perfect CSI due to the high mobility of UAVs.

Since RTNA outperforms Semi-RTNA in terms of network ST, we then verify
the efficiency of the PAE policy in improving network ST under RTNA.
In particular, we study the scaling behavior of network ST with the
growing UAV deployment density.

\begin{theorem}

When each UAV adjusts the half-beamwidth according to $\Phi^{*}=\arctan\left(\frac{C}{\Delta h\sqrt{\lambda_{\mathrm{a}}}}\right)$,
the scaling behavior of network ST under RTNA rule is given by
\begin{align}
\underset{\lambda\rightarrow\infty}{\mathrm{lim}} & \mathsf{ST}_{\mathrm{dir}}^{\mathrm{R}}=\lambda_{\mathrm{a}}\hat{p}_{\mathrm{p}}^{\dagger}\delta_{1}\left(\hat{p}_{\mathrm{p}}^{\dagger},C,\alpha,\tau\right)\min\left\{ \log_{2}\left(1+\tau\right),C_{\mathrm{b}}\right\} ,\label{eq:ST scaling law}
\end{align}
where $\hat{p}_{\mathrm{p}}^{\dagger}=1-\exp\left(-\pi C^{2}\right)$
and $\delta_{1}\left(\hat{p}_{\mathrm{p}}^{\dagger},C,\alpha,\tau\right)=\frac{\exp\left(-\hat{p}_{\mathrm{p}}^{\dagger}\pi C^{2}\omega_{2}\left(\alpha,\frac{1}{\tau}\right)\right)-\exp\left(\pi C^{2}\right)}{1-\hat{p}_{\mathrm{p}}^{\dagger}\omega_{2}\left(\alpha,\frac{1}{\tau}\right)}$.

\label{theorem: ST scaling behavior}

\end{theorem}

\textit{Proof}: Please refer to Appendix \ref{subsec:Proof ST scaling behavior}.\qed

According to Theorem \ref{theorem: ST scaling behavior}, we further
study the network ST scaling behavior in the following two cases.

\textbf{1) Backhaul unlimited case.} This case occurs when sufficient
backhaul could be provided through mmWave and GU density is limited.
Accordingly, (\ref{eq:ST scaling law}) in Theorem \ref{theorem: ST scaling behavior}
degenerates into
\begin{align}
\underset{\lambda\rightarrow\infty}{\mathrm{lim}}\mathsf{ST}_{\mathrm{dir}}^{\mathrm{R}} & \overset{\left(\mathrm{a}\right)}{=}\lambda_{\mathrm{GU}}\hat{p}_{\mathrm{p}}^{\dagger}\delta_{1}\left(\hat{p}_{\mathrm{p}}^{\dagger},C,\alpha,\tau\right)\log_{2}\left(1+\tau\right),\label{eq:ST scaling law backhaul unlimited}
\end{align}
where (a) follows because $\lambda_{\mathrm{a}}=q_{\mathrm{a}}\lambda$
and $\underset{\lambda\rightarrow\infty}{\mathrm{lim}}q_{\mathrm{a}}=\lambda_{\mathrm{GU}}/\lambda$
according to (\ref{eq:activation probability}). It is shown in (\ref{eq:ST scaling law backhaul unlimited})
that network ST will linearly increase with the GU density $\lambda_{\mathrm{GU}}$.
This indicates that greater network ST could be obtained if more GUs
request service from UAVs. In other words, the proposed PAE policy
could sustainably improve spectrum reuse as long as sufficient wireless
backhaul is provided. This is fundamentally different from the cases
in Figs. \ref{fig:comparison of semi-RTNA and RTNA} and \ref{fig:Impact of beamwidth b},
where network ST is degraded by over-deployment of UAVs supposing
that constant-beamwidth directional-antenna is equipped by each UAV.

More importantly, it is shown in (\ref{eq:ST scaling law backhaul unlimited})
that identical network ST can be obtained even if UAVs are in different
altitudes with the application of the proposed PAE policy. For this
reason, the optimization of beamwidth could help compensate for the
loss of network ST, which is due to the increase flight altitude.

\textbf{2) Backhaul limited case.} This case occurs when either limited
backhaul could be provided or GU density is sufficiently large. According
to Theorem \ref{theorem: ST scaling behavior}, (\ref{eq:ST scaling law})
degenerates into
\begin{align}
\underset{\lambda\rightarrow\infty}{\mathrm{lim}}\mathsf{ST}_{\mathrm{dir}}^{\mathrm{R}} & \overset{\left(\mathrm{a}\right)}{=}\frac{\hat{p}_{\mathrm{p}}^{\dagger}\delta_{1}\left(\hat{p}_{\mathrm{p}}^{\dagger},C,\alpha,\tau\right)G_{\mathrm{m}}\left(\vartheta\right)F_{\left|\varepsilon\right|}\left(\vartheta\right)C_{\mathrm{t}}}{\pi R_{\mathrm{m}}^{2}}.\label{eq:ST scaling law backhaul limited}
\end{align}
In this case, it is observed from (\ref{eq:ST scaling law backhaul limited})
that network ST is limited by the mmWave link capacity and mmWave
effective function, which is dependent on the UAV flight altitude
and hover radius (see Section \ref{subsec:Backhaul-Model}). According
to Lemma \ref{lemma: error function}, $G_{\mathrm{m}}\left(\vartheta\right)F_{\left|\varepsilon\right|}\left(\vartheta\right)$
is a decreasing function of $\vartheta=2\arctan\left(\frac{R_{\mathrm{h}}}{h_{\mathrm{UAV}}}\right)$.
Therefore, the convergence value of network ST can be improved by
increasing the flight altitude under the PAE policy in the backhaul-limited
case.

\begin{figure}[t]
\centering{}\includegraphics[width=3.5in]{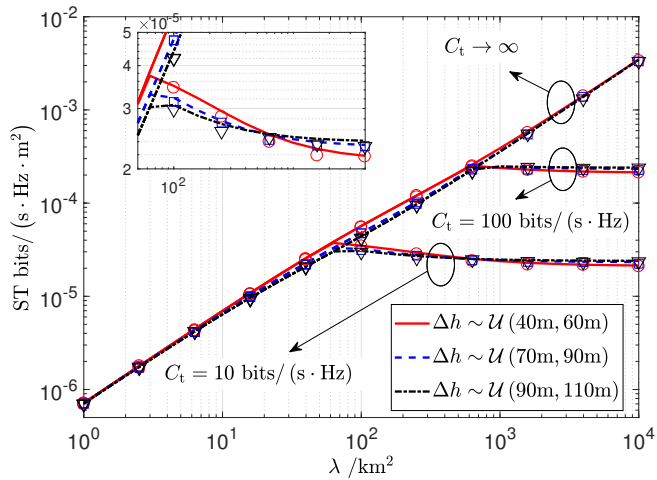}\caption{\label{fig:Impact of PAE}Network ST v.s. UAV density under PAE policy
with projection scaling parameter $C=1$. Note that RTNA is applied
for user association. Except for $\Delta h$, other system parameters
are identically set as those in Fig. \ref{fig:Impact of hover radius}.}
\end{figure}

Finally, we verify the efficiency of the proposed beamwidth optimization
policy. In particular, Fig. \ref{fig:Impact of PAE} plots network
ST as a function of UAV density under RTNA. Compared to the results
in Fig. \ref{fig:Impact of beamwidth b}, network ST could linearly
increase and then converge with the growing UAV density when RTNA
is applied under the PAE policy. Moreover, even if UAVs are deployed
over different altitudes, identical network ST could be obtained supposing
that the backhaul capacity is sufficiently large, i.e., $C_{\mathrm{t}}\rightarrow\infty$,
which verifies the validity of the results in (\ref{eq:ST scaling law backhaul unlimited}).
Otherwise, when backhaul capacity is limited (see the lines under
$C_{\mathrm{t}}=10\mathrm{bits/\left(\mathrm{s}\cdot\mathrm{Hz}\right)}$
or $C_{\mathrm{t}}=100\mathrm{bits/\left(\mathrm{s}\cdot\mathrm{Hz}\right)}$
in Fig. \ref{fig:Impact of PAE}), increasing the flight altitude
could enhance network ST under large UAV deployment density.

In practice, the UAV flight altitude has to be raised in specific
scenarios such as emergency and cooperative engagement, which is detrimental
to the spatial reuse of available spectrum resources. However, the
proposed simple but effective PAE policy could significantly enhance
the spatial reuse in higher flight altitude considering both backhaul
unlimited and backhaul-unlimited cases. Moreover, it is worth noting
that the PAE policy could be feasibly applied in the multi-antenna
system \cite{Ref_MIMO_extension,Ref_MIMO_extension_2}, where UAV
APs are equipped with multiple antennas. The principle of PAE and
multi-antenna techniques such as coordinated beamforming is to enhance
the desired signal power and alleviate the interference power received
at the intended receiver. The difference is that received power distribution
will be different and accordingly the UAV beamwidth should be further
optimized, which is dependent on the applied multi-antenna techniques.

\section{Conclusion\label{sec:Conclusion}}

In this paper, we have modeled and evaluated the performance of fixed-wing
UAV network, where UAVs serve as APs to provide air-to-ground coverage
to GUs with mmWave backhaul. It was shown that the hovering feature
of UAV would result in unstable connections of GUs and reduce the
mmWave mainlobe gain, which degrades CP and network ST. Even though
the impact of UAV hovering is minor supposing that GUs are associated
with the closest UAVs in a real-time manner, excessive handover overhead
will be introduced. More importantly, it was shown that the over-deployment
of UAVs would result in a rapid decrease of network ST since coverage
of aerial APs is large. To enhance the performance of the fixed-wing
UAV network, we propose a PAE policy to adaptively adjust UAV beamwidth
according to the UAV deployment density. Notably, network was proved
to increase with the UAV density and independent of the UAV flight
altitude under PAE. Therefore, the results of this paper are helpful
for the design, deployment and beamwidth optimization of fixed-wing
UAV network.

\appendix

\section{*}

\subsection{Proof for Lemma \ref{lemma: error function}\label{subsec:Proof error function}}

According to \cite{Ref_mmWave_model}, we assume that the absolute
mmWave beam orientation error $\left|\varepsilon\right|$ follows
exponential distribution, which is truncated to $\left[0,\pi\right]$.
Therefore, the PDF of $\left|\varepsilon\right|$ is given by 

\begin{align}
f_{\left|\varepsilon\right|}\left(x\right) & =\begin{cases}
\frac{\mathrm{exp}\left(-\frac{x}{\bar{\varepsilon}}\right)}{\left(1-\mathrm{exp}\left(-\frac{\pi}{\bar{\varepsilon}}\right)\right)\bar{\varepsilon}}, & x\in\left[0,\pi\right]\\
0. & \mathrm{otherwise}
\end{cases}\label{eq:PDF of error}
\end{align}
Since mmWave backhaul is effective only if $\left|\varepsilon\right|\leq\frac{\vartheta}{2}$,
the proof can be completed by computing $\intop_{0}^{\vartheta/2}f_{\left|\varepsilon\right|}\left(x\right)\mathrm{d}x$.

\subsection{Proof for Proposition \ref{proposition: CP and ST Semi-RTNA omni}\label{subsec:Proof CP ST semi-RTNA omni}}

Since each UAV independently hovers with identical radius \textcolor{black}{$R_{\mathrm{h}}$,}
hover angle $\theta$ follows uniform distribution over $\left[0,2\pi\right]$,
i.e., $\theta\sim\mathcal{U}\left(0,2\pi\right)$. In consequence,
the 2D locations of UAVs still follow PPP with density $\lambda$
\cite{Ref_PPP_feature}. Meanwhile, when the half-beamwidth of UAV
directional antenna equals $\frac{\pi}{2}$, GUs are always in the
projection area of the associated UAVs, i.e., $p_{\mathrm{p}}=1$.
Therefore, the CP under the Semi-RTNA is given by

\begin{align}
\mathsf{CP}_{\pi/2}^{\mathrm{SR}}= & \mathbb{P}\left\{ \frac{PH_{\mathrm{UAV}_{0}}d_{0}^{-\alpha}}{I^{\mathrm{SR}}}>\tau\right\} \nonumber \\
= & \mathbb{P}\left\{ H_{\mathrm{UAV}_{0}}>\frac{\tau I^{\mathrm{SR}}}{Pd_{0}^{-\alpha}}\right\} \nonumber \\
= & \mathbb{E}_{r_{0},\theta,\Delta h}\left[\mathcal{L_{\mathit{I}^{\mathrm{SR}}}}\left(s_{1}\right)\right],\label{eq:CP semi-RTNA omni - proof 1}
\end{align}
where $\mathcal{L_{\mathit{I}^{\mathrm{SR}}}}\left(s_{1}\right)=\exp\left(-s_{1}I^{\mathrm{SR}}\right)$
denotes the Laplace Transform of $I^{\mathrm{SR}}$ evaluated at $s_{1}=\frac{\tau}{Pd_{0}^{-\alpha}}$.
Note that $d_{0}=\sqrt{r_{0}^{2}+R_{\mathrm{h}}^{2}-2R_{\mathrm{h}}r_{0}\cos\theta+\Delta h^{2}}$
follows according to Law of Cosines \cite{Ref_cosine_law}. In particular,
$\mathcal{L_{\mathit{I}^{\mathrm{SR}}}}\left(s_{1}\right)$ can be
expressed as

\begin{align}
\mathcal{L_{\mathit{I}^{\mathrm{SR}}}}\left(s_{1}\right)\overset{\left(\mathrm{a}\right)}{=} & \exp\left[-2\pi\lambda_{\mathrm{a}}\int_{l}^{\infty}x\left(1-\frac{1}{1+s_{1}Px^{-\alpha}}\right)\mathrm{d}x\right]\nonumber \\
= & \exp\left[-2\pi\lambda_{\mathrm{a}}\frac{\tau d_{0}^{\alpha}l^{2-\alpha}\omega_{1}\left(\alpha,\tau d_{0}^{\alpha}l^{-\alpha}\right)}{\alpha-2}\right],\label{eq:CP semi-RTNA omni - proof 2}
\end{align}
where (a) follows owing to the probability generating functional (PFGL)
of PPP \cite{book_stochastic_geometry}. Given $r_{0}\leq R_{\mathrm{h}}$,
the hovering interfering UAVs can be sufficiently close to $\mathrm{GU}_{0}$
such that $l=\Delta h$. Otherwise, when $r_{0}\geq R_{\mathrm{h}}$,
we have $l=\sqrt{\left(r_{0}-R_{\mathrm{h}}\right)^{2}+\Delta h^{2}}$. 

In addition, the PDF of $r_{0}$ can be obtained using contact distribution
\cite{book_stochastic_geometry}, i.e., $f_{r_{0}}\left(x\right)=2\pi\lambda x\exp\left(-\pi\lambda x^{2}\right),x\in\left(0,\infty\right)$.
Hence, CP and network ST can be obtained through substituting (\ref{eq:CP semi-RTNA omni - proof 2})
into (\ref{eq:CP semi-RTNA omni - proof 1}) and deconditioning $r_{0}$,
$\theta$ and $\Delta h$.

\subsection{Proof for Corollary \ref{corollary: CP and ST Semi-RTNA omni upper bound}\label{subsec:Proof CP ST semi-RTNA omni UB}}

According to (\ref{eq:CP semi-RTNA omni}) in Proposition \ref{proposition: CP and ST Semi-RTNA omni},
when UAV flight altitude is fixed, the conditional CP is given by
\begin{align}
\mathsf{CP}_{\pi/2\left|\Delta h\right.}^{\mathrm{SR}} & =\mathbb{E}_{r_{0},\theta}\left[\exp\left(-2\pi\lambda_{\mathrm{a}}\frac{\tau d_{0}^{\alpha}l^{2-\alpha}\omega_{1}\left(\alpha,\tau d_{0}^{\alpha}l^{-\alpha}\right)}{\alpha-2}\right)\right]\nonumber \\
 & \overset{\left(\mathrm{a}\right)}{<}\mathbb{E}_{r_{0},\theta}\left[\exp\left(-2\pi\lambda_{\mathrm{a}}\frac{\tau d_{0}^{\alpha}\hat{d}_{0}^{2-\alpha}\omega_{1}\left(\alpha,\tau d_{0}^{\alpha}l^{-\alpha}\right)}{\alpha-2}\right)\right]\nonumber \\
 & \overset{\left(\mathrm{b}\right)}{<}\mathbb{E}_{r_{0},\theta}\left[\exp\left(-2\pi\lambda_{\mathrm{a}}\frac{\tau\hat{d}_{0}^{2}\omega_{1}\left(\alpha,\tau d_{0}^{\alpha}l^{-\alpha}\right)}{\alpha-2}\right)\right]\nonumber \\
 & \overset{\left(\mathrm{c}\right)}{<}\mathbb{E}_{r_{0}}\left[\exp\left(-\pi\lambda\delta\left(\alpha,\tau,q_{\mathrm{a}}\right)\hat{d}_{0}^{2}\right)\right]\nonumber \\
 & =\exp\left(-\pi\lambda\delta\left(\alpha,\tau,q_{\mathrm{a}}\right)\Delta h^{2}\right)\mathbb{E}_{r_{0}}\left[\exp\left(-\pi\lambda\delta\left(\alpha,\tau,q_{\mathrm{a}}\right)\left(r_{0}-R_{\mathrm{h}}\right)^{2}\right)\right]\nonumber \\
 & =\hat{\mathsf{CP}}_{\pi/2\left|\Delta h\right.}^{\mathrm{SR}}\left(\lambda\right),\label{eq:CP semi-RTNA omni UB proof}
\end{align}
where (a) follows because we substitute $l$ by $\hat{d}_{0}=\sqrt{\left(r_{0}-R_{\mathrm{h}}\right)^{2}+\Delta h^{2}}$
and $\hat{d}_{0}>l$, (b) follows because $\hat{d}_{0}<d_{0}=\sqrt{r_{0}^{2}+R_{\mathrm{h}}^{2}-2r_{0}R_{\mathrm{h}}\cos\theta+\Delta h^{2}}$,
and (c) follows because we substitute $l$ by $d_{0}$, $l<d_{0}$
and $\omega_{1}\left(x,y\right)$ is a decreasing function of $x$
(see Lemma 1 in \cite{Ref_SBPM}).

\subsection{Proof for Proposition \ref{proposition: CP and ST RTNA omni}\label{subsec:Proof CP ST RTNA omni}}

Similarly to the proof in Appendix \ref{subsec:Proof CP ST semi-RTNA omni},
$\mathsf{CP}_{\pi/2}^{\mathrm{R}}$ under RTNA is given by

\begin{align}
\mathsf{CP}_{\pi/2}^{\mathrm{R}}= & \mathbb{E}_{r_{0},\Delta h}\left[\mathcal{L_{\mathit{I}^{\mathrm{R}}}}\left(s_{2}\right)\right],\label{eq:CP RTNA omni - proof 1}
\end{align}
where $s_{2}=\frac{\tau}{P\check{d}_{0}^{-\alpha}}$ and $\check{d}_{0}=\sqrt{r_{0}^{2}+\Delta h^{2}}$.
The Laplace Transform $\mathcal{L_{\mathit{I}^{\mathrm{R}}}}\left(s_{2}\right)$
can be obtained by

\begin{align}
\mathcal{L_{\mathit{I}^{\mathrm{R}}}}\left(s_{2}\right)= & \mathbb{E}_{\mathit{I}^{\mathrm{R}}}\left[e^{-s_{2}\mathit{I}^{\mathrm{R}}}\right]\nonumber \\
= & \mathbb{E_{\mathrm{\mathit{I}^{\mathrm{R}}}}}\left[\prod\frac{1}{1+s_{2}P\check{d}_{0}^{-\alpha}}\right]\nonumber \\
\overset{\left(\mathrm{a}\right)}{=} & \exp\left(-2\pi\lambda_{\mathrm{a}}\int_{\check{d}_{0}}^{\infty}x\left(1-\frac{1}{1+s_{2}Px^{-\alpha}}\right)\mathrm{d}x\right)\nonumber \\
= & \exp\left(-2\pi\lambda_{\mathrm{a}}\frac{\tau\check{d}_{0}^{2}{}_{2}F_{1}\left(1,1-\frac{2}{\alpha},2-\frac{2}{\alpha},-\tau\right)}{\alpha-2}\right)\nonumber \\
\overset{\left(\mathrm{b}\right)}{=} & \exp\left(-\pi\lambda\delta\left(\alpha,\tau,q_{\mathrm{a}}\right)r_{0}^{2}\right)\exp\left(-\pi\lambda\delta\left(\alpha,\tau,q_{\mathrm{a}}\right)\Delta h^{2}\right),\label{eq:CP RTNA omni - proof 2}
\end{align}
where $\delta\left(\alpha,\tau,q_{\mathrm{a}}\right)=\frac{2q_{\mathrm{a}}\tau\omega_{1}\left(\alpha,\tau\right)}{\alpha-2}$.
In (\ref{eq:CP RTNA omni - proof 2}), (a) follows due to probability
generating functional (PFGL) of PPP \cite{book_stochastic_geometry}
and (b) follows due to $\check{d}_{0}=\sqrt{r_{0}^{2}+\Delta h^{2}}$
since GUs always connect to the closest UAVs for service under RTNA.
In this case, the CP of GUs is independent of the hover angle.

With (\ref{eq:CP RTNA omni - proof 1}), (\ref{eq:CP RTNA omni - proof 2})
and the \textcolor{black}{uniform distribution $\Delta h\sim\mathcal{U}\left(\tilde{h}_{\mathrm{L}},\tilde{h}_{\mathrm{U}}\right)$,
we obtain the }$\mathsf{CP}_{\pi/2}^{\mathrm{R}}$ by deconditioning
$\Delta h$ as

\begin{align}
\mathsf{CP}_{\pi/2}^{\mathrm{R}}= & \mathbb{E}_{\Delta h}\left[\frac{\exp\left(-\pi\lambda\delta\left(\alpha,\tau,q_{\mathrm{a}}\right)\Delta h^{2}\right)}{1+\delta\left(\alpha,\tau,q_{\mathrm{a}}\right)}\right]\nonumber \\
= & \frac{\mathrm{Erf}\left(\sqrt{\delta\left(\alpha,\tau,q_{\mathrm{a}}\right)}\tilde{h}_{\mathrm{U}}\right)-\mathrm{Erf}\left(\sqrt{\delta\left(\alpha,\tau,q_{\mathrm{a}}\right)}\tilde{h}_{\mathrm{L}}\right)}{2\sqrt{\lambda\delta\left(\alpha,\tau,q_{\mathrm{a}}\right)}\left(1+\delta\left(\alpha,\tau,q_{\mathrm{a}}\right)\right)\left(\tilde{h}_{\mathrm{U}}-\tilde{h}_{\mathrm{L}}\right)}.
\end{align}

\subsection{Proof for Proposition \ref{proposition: CP and ST Semi-RTNA dir}\label{subsec:Proof CP ST semi-RTNA dir}}

When directional-antenna is equipped by each UAV under Semi-RTNA,
$\mathsf{CP}_{\mathrm{dir}\left|r_{0},\theta,\Delta h\right.}^{\mathrm{SR}}$
is given by

\begin{align}
\mathsf{CP}_{\mathrm{dir}\left|r_{0},\theta,\Delta h\right.}^{\mathrm{SR}}=p_{\mathrm{p}} & \mathbb{E}_{r_{0},\theta,\Delta h}\left[\mathcal{L_{\hat{I}^{\mathrm{SR}}}}\left(s_{3}\right)\right],\label{eq:CP semi-RTNA dir - proof 1}
\end{align}
where $\mathcal{L_{\hat{I}^{\mathrm{SR}}}}\left(s_{3}\right)$ denotes
the Laplace Transform of $\hat{I}^{\mathrm{SR}}$ at $s_{3}=\frac{\tau}{PG\left(\phi,\varphi\right)d_{0}^{-\alpha}}$.
With the projection probability $p_{\mathrm{p}}$ given by (\ref{eq:projection probability}),
$\mathcal{L_{\hat{I}^{\mathrm{SR}}}}\left(s_{3}\right)$ can be obtained
as

\begin{align}
\mathcal{L_{\hat{I}^{\mathrm{SR}}}}\left(s_{3}\right)= & \mathbb{E}_{\hat{I}^{\mathrm{SR}}}\left[\exp\left(-s_{3}\hat{I}^{\mathrm{SR}}\right)\right]\nonumber \\
= & \mathbb{E_{\hat{\mathit{I}}^{\mathrm{SR}}}}\left[\prod\frac{1}{1+s_{3}PG\left(\phi,\varphi\right)d_{0}^{-\alpha}}\right]\nonumber \\
\overset{\left(\mathrm{a}\right)}{=} & \exp\left(-2\pi p_{\mathrm{p}}\lambda_{\mathrm{a}}\int_{\hat{l}}^{\hat{R}_{\mathrm{p}}}x\left(1-\frac{1}{1+s_{3}PG\left(\phi,\varphi\right)x^{-\alpha}}\right)\mathrm{d}x\right)\nonumber \\
= & \exp\left(-\pi p_{\mathrm{p}}\lambda_{\mathrm{a}}\left(\hat{R}_{\mathrm{p}}^{2}\omega_{2}\left(\alpha,\frac{\hat{R}_{\mathrm{p}}^{\alpha}}{\tau d_{0}^{\alpha}}\right)-\hat{l}^{2}\omega_{2}\left(\alpha,\frac{\hat{l}^{\alpha}}{\tau d_{0}^{\alpha}}\right)\right)\right),\label{eq:CP semi-RTNA dir - proof 2}
\end{align}
where (a) follows since the inter-cell interference is limited due
to the projection of directional-antenna UAVs. Specifically, interference
only stems from the set of UAVs with density $p_{\mathrm{p}}\lambda_{\mathrm{a}}$,
which are distributed within the projection area around $\mathrm{GU}_{0}$
with radius $R_{\mathrm{p}}$.

In (\ref{eq:CP semi-RTNA dir - proof 2}), when\textcolor{black}{{}
$r_{0}\leq R_{\mathrm{h}}\leq R_{\mathrm{p}}$ holds, the }distance
from the interfering UAVs to $\mathrm{GU}_{0}$ is greater than $\hat{l}=\Delta h$.
Otherwise, when \textcolor{black}{$r_{0}>R_{\mathrm{h}}$,} \textcolor{black}{the}
distance from the closest interfering UAVs to $\mathrm{GU}_{0}$ is
$\hat{l}=\sqrt{\left(r_{0}-R_{\mathrm{h}}\right)^{2}+\Delta h^{2}}$.

Integrating (\ref{eq:CP semi-RTNA dir - proof 1}) and (\ref{eq:CP semi-RTNA dir - proof 2}),
we complete the proof.

\subsection{Proof for Lemma \ref{lemma: increasing function} \label{subsec:Proof increasing function}}

The proof is completed by showing the derivative of $\Psi\left(x\right)$
in terms of $x$ is greater than 0. Specifically, the derivative is
given by
\begin{align}
\text{\ensuremath{\frac{\mathrm{d}\Psi\left(x\right)}{\mathrm{d}x}}=} & \frac{2x}{1+bx^{z}}.\label{eq:increasing function proof}
\end{align}
Since $b>0$, $\frac{\mathrm{d}\Psi\left(x\right)}{\mathrm{d}x}>0$
holds.

\subsection{Proof for Proposition \ref{proposition: CP and ST RTNA dir}\label{subsec:Proof CP ST RTNA dir}}

When directional-antenna is equipped by each UAV under RTNA, we have
$\hat{p}_{\mathrm{p}}=1-\exp\left(\pi\lambda R_{\mathrm{p}}^{2}\right)$.
Therefore, $\mathsf{CP}_{\mathrm{dir}}^{\mathrm{R}}$ is given by

\begin{align}
\mathsf{CP}_{\mathrm{dir}}^{\mathrm{R}} & =\mathbb{E}_{r_{0},\Delta h}\left[\hat{p}_{\mathrm{p}}\mathcal{L_{\hat{I}^{\mathrm{R}}}}\left(s_{4}\right)\right]\nonumber \\
 & =\mathbb{E}_{r_{0},\Delta h,\hat{I}^{\mathrm{R}}}\left[\hat{p}_{\mathrm{p}}\exp\left(-s_{4}\hat{I}^{\mathrm{R}}\right)\right]\nonumber \\
 & =\mathbb{E}_{r_{0},\Delta h}\left[\hat{p}_{\mathrm{p}}\exp\left(-2\pi\hat{p}_{\mathrm{p}}\lambda_{\mathrm{a}}\int_{\check{d}_{0}}^{\hat{R}_{\mathrm{p}}}x\left(1-\frac{1}{1+s_{4}PG\left(\phi,\varphi\right)x^{-\alpha}}\right)\mathrm{d}x\right)\right]\nonumber \\
 & =\mathbb{E}_{r_{0},\Delta h}\left[\hat{p}_{\mathrm{p}}\exp\left[-\pi\hat{p}_{\mathrm{p}}\lambda_{\mathrm{a}}\left(\hat{R}_{\mathrm{p}}^{2}\omega_{2}\left(\alpha,\frac{\hat{R}_{\mathrm{p}}^{\alpha}}{\tau\check{d}_{0}^{\alpha}}\right)-\check{d}_{0}^{2}\omega_{2}\left(\alpha,\frac{1}{\tau}\right)\right)\right]\right],
\end{align}
where $\mathcal{L_{\hat{I}^{\mathrm{R}}}}\left(s_{4}\right)$ denotes
the Laplace Transform of $\hat{I}^{\mathrm{R}}$ at $s_{4}=\frac{\tau}{PG\left(\phi,\varphi\right)\check{d}_{0}^{-\alpha}}$,
$\check{d}_{0}=\sqrt{r_{0}^{2}+\Delta h^{2}}$ and $\hat{I}^{\mathrm{R}}=\underset{\tiny{\mathrm{UAV}_{i}\in\mathrm{\tilde{\Pi}_{UAV}}\backslash\mathrm{UAV}_{0}}}{\sum}PH_{\mathrm{UAV}_{i}}G\left(\phi,\varphi\right)d_{i}^{-\alpha}$
is the interference stemming from the overlapping activated UAV cells.
The remaining parts of the proof are identical to those in Appendix
\ref{subsec:Proof CP ST RTNA omni} and thus omitted due to space
limitation.

\subsection{Proof for Theorem \ref{theorem: ST scaling behavior}\label{subsec:Proof ST scaling behavior}}

According to Proposition \ref{proposition: CP and ST RTNA dir}, given
a fixed flight altitude, the conditional CP of GUs is given by
\begin{align}
\underset{\lambda\rightarrow\infty}{\mathrm{lim}}\mathsf{CP}_{\mathrm{dir}\left|\Delta h\right.}^{\mathrm{R}}\left(\lambda\right)= & \hat{p}_{\mathrm{p}}\mathbb{E}_{r_{0}<R_{\mathrm{p}}}\left[\exp\left(-\pi\hat{p}_{\mathrm{p}}\lambda_{\mathrm{a}}\left(\hat{R}_{\mathrm{p}}^{2}\omega_{2}\left(\alpha,\frac{\hat{R}_{\mathrm{p}}^{\alpha}}{\tau\check{d}_{0}^{\alpha}}\right)-\check{d}_{0}^{2}\omega_{2}\left(\alpha,\frac{1}{\tau}\right)\right)\right)\right]\nonumber \\
\overset{\left(\mathrm{a}\right)}{=} & \hat{p}_{\mathrm{p}}\mathbb{E}_{r_{0}<R_{\mathrm{p}}}\left[\exp\left(-\pi\hat{p}_{\mathrm{p}}\lambda_{\mathrm{a}}\omega_{2}\left(\alpha,\frac{1}{\tau}\right)\left(R_{\mathrm{p}}^{2}-r_{0}^{2}\right)\right)\right]\nonumber \\
= & \hat{p}_{\mathrm{p}}\frac{\exp\left(-\hat{p}_{\mathrm{p}}\pi\lambda_{\mathrm{a}}R_{\mathrm{p}}^{2}\omega_{2}\left(\alpha,\frac{1}{\tau}\right)\right)-\exp\left(-\pi\lambda_{\mathrm{a}}R_{\mathrm{p}}^{2}\right)}{1-\hat{p}_{\mathrm{p}}\omega_{2}\left(\alpha,\frac{1}{\tau}\right)},\label{eq:proof ST limiting behavior 1}
\end{align}
where (a) follows because $\underset{\lambda\rightarrow\infty}{\lim}R_{\mathrm{p}}\rightarrow0$
and $\underset{\lambda\rightarrow\infty}{\lim}r_{0}\rightarrow0$.
Substituting $R_{\mathrm{p}}=\frac{C}{\sqrt{\lambda_{\mathrm{a}}}}$
into (\ref{eq:proof ST limiting behavior 1}), we have
\begin{align}
\underset{\lambda\rightarrow\infty}{\mathrm{lim}}\mathsf{CP}_{\mathrm{dir}\left|\Delta h\right.}^{\mathrm{R}}\left(\lambda\right) & =\hat{p}_{\mathrm{p}}^{\dagger}\frac{\exp\left(-\hat{p}_{\mathrm{p}}^{\dagger}\pi C^{2}\omega_{2}\left(\alpha,\frac{1}{\tau}\right)\right)-\exp\left(\pi C^{2}\right)}{1-\hat{p}_{\mathrm{p}}^{\dagger}\omega_{2}\left(\alpha,\frac{1}{\tau}\right)}.\label{eq:proof ST limiting behavior 2}
\end{align}
In (\ref{eq:proof ST limiting behavior 2}), $\underset{\lambda\rightarrow\infty}{\mathrm{lim}}\mathsf{CP}_{\mathrm{dir}\left|\Delta h\right.}^{\mathrm{R}}\left(\lambda\right)$
is independent of UAV flight altitude. Therefore, $\underset{\lambda\rightarrow\infty}{\mathrm{lim}}\mathsf{CP}_{\mathrm{dir}}^{\mathrm{R}}\left(\lambda\right)=\underset{\lambda\rightarrow\infty}{\mathrm{lim}}\mathsf{CP}_{\mathrm{dir}\left|\Delta h\right.}^{\mathrm{R}}\left(\lambda\right)$.
Hence, we complete the proof.

\bibliographystyle{IEEEtran}
\bibliography{ref_UAV}

\end{document}